\documentclass[a4paper,11pt]{article}
\usepackage{jheppub} 
\usepackage{lineno}
\usepackage{mathrsfs}
\usepackage{bbm}
\usepackage{mathabx}

\arxivnumber{2509.01949} 

\title{\boldmath Free-field approaches to boundary $\mathcal{W} \big[ \widehat{g} \big] (p,p') $ minimal models}

\author{Xun Liu}
\affiliation{Department of Physics, The University of Tokyo \\
7-3-1 Hongo, Bunkyo-ku, Tokyo 113-0033, Japan}

\emailAdd{xun.liu@tnp.phys.s.u-tokyo.ac.jp}
\abstract{We apply the background charged bosonic free-field approach to the rational principal quantum Drinfeld-Sokolov (QDS) $\mathcal{W} \big[ \widehat{g} \big](p,p')$ minimal models with boundaries, where $g$ is a finite bosonic simple Lie algebra. Their Ishibashi states are expressed using the free bosonic Ishibashi states, by applying the Fock space resolutions. The Coulomb-gas formalism is applied to the calculations of the disk two-point correlation functions in some well-studied and lesser familiar rational QDS $\mathcal{W} \big[ \widehat{g} \big](p,p')$ minimal models. The analytical expressions can be obtained by the repeated applications of the Pochhammer contour integral expression and the Taylor expansions of Lauricella's hypergeometric functions $F_{D}^{(n)}$. }

\begin{document}
\maketitle
\flushbottom

\section{Introduction}
\label{sec:intro}

Two-dimensional conformal field theories (CFT$_2$s) are quantum field theories (QFTs) that are invariant under two-dimensional conformal transformations. The two-dimensional local conformal symmetries are described by infinite-dimensional local conformal algebras $\mathcal{A}\otimes\widebar{\mathcal{A}} $. $\mathcal{A}$ and $\widebar{\mathcal{A}}$ are infinite-dimensional Lie algebras, called the chiral and anti-chiral algebras \cite{Belavin:1984vu, Moore:1988qv}. $\mathcal{A}$ and $\widebar{\mathcal{A}}$ are not necessarily identical. However, in this work, only the cases with $\mathcal{A}= \widebar{\mathcal{A}}$ are considered. The infinite-dimensional local conformal symmetries provide us with stronger analytical controls over the physical observations of CFT$_2$s, such as the partition functions and the correlation functions \cite{Belavin:1984vu, Cardy:1986ie, Moore:1988qv}, allowing us to solve those quantities in some of the CFT$_2$ analytically.

In this work, we focus on rational conformal field theories (RCFT$_2$s) that admit free-field approaches. RCFT$_2$s are the simplest class of CFT$_2$s, where the fusion algebra of the irreducible representations of the theory is finite and closed \cite{Anderson:1987ge, Moore:1988qv}. We claim that a CFT$_2$ admits a free-field approach if the following three conditions are satisfied \cite{Bouwknegt:1990wa}. 
\begin{itemize}
    \item  The chiral algebra $\mathcal{A}$ and the anti-chiral algebra $\widebar{\mathcal{A}}$ can be realized using the free-fields. 

    \item  The existence of projection maps from the free-field Fock space modules to chiral modules in the original CFT$_2$. These projections maps are referred to as the Fock space resolutions of the chiral modules. 

    \item  The validity of the Coulomb-gas formalism, that is, the correlation functions of the original CFT$_2$ can be calculated using the free-field vertex operators and the free-field $\mathcal{A}$ ($\widebar{\mathcal{A}}$) intertwiners.
\end{itemize}  
There is a wide range of RCFTs that are known to admit free-field approaches. Two canonical archetypes are
\begin{itemize}
    \item The positive integral level Kac-Moody (KM) theories, or more often known as the rational Wess-Zumino-Witten (WZW) models  \cite{Witten:1983ar, Knizhnik:1984nr, Gepner:1986wi}. A free-field approach to them is the Wakimoto-type free approach, realized from background-charged bosonic fields $\phi$ and bosonic $\beta \gamma$ ghost systems \cite{Wakimoto:1986gf, Bernard:1989iy, Bouwknegt:1989xa, Bouwknegt:1989jf, Bouwknegt:1990aa, Bouwknegt:1990wa, Bouwknegt:1991gf}. 

    \item  The principal quantum Drinfeld-Sokolov (QDS) $\mathcal{W}$ minimal models. A background-charged bosonic free-field approach to the is induced from the Wakimoto-type of free approach to KM theories \cite{Dotsenko:1984nm, Dotsenko:1984ad, Dotsenko:1985hi, Fateev:1987vh, Fateev:1987zh, Felder:1988zp, Bershadsky:1989mf, Bouwknegt:1992wg}.
\end{itemize}

\

We apply the background-charged free-field approach to rational principal QDS $\mathcal{W}\big[ \widehat{g} \big] (p,p')$ minimal models ($g$ being simple Lie algebras). Their Ishibashi states are expressed as infinite linear combinations of the Fock space Ishibashi states. And the Coulomb-gas formalism is applied to the calculations of the disk bulk two-point functions \cite{Dotsenko:1984nm, Dotsenko:1984ad, Dotsenko:1985hi}, in both well-studied and less familiar models. The results of the Coulomb-gas integral can be obtained analytically by the repeated applications of the Pochhamer contour integral expression and the Taylor expansions of the Lauricella hypergeometric functions $F_{D}^{(n)}$. Note that, we avoid the Pochhammer contour deformation to the real axis segment integrals, since the Pochhammer contour integrals are always finite, while the real axis segment integrals can be divergent.


This work is organized as following. In Chapter \ref{sec: rev}, the necessary knowledge of finite Lie algebras $g$, their related Kac-Moody algebras $\widehat{g}$, the definition of the principal QDS $\mathcal{W} \big[ \widehat{g} \big]$ algebras, the free-field approaches to them, and the concepts in boundary RCFTs. In Chapter \ref{sec: Uni re}, model-independent results are described, including the free-field expressions of the $\mathcal{W}$ minimal Ishibashi states and their disk one-point functions. In Chapter \ref{Sec: Disk 2pt}, disk correlation bulk two-point functions in some Virasoro, $\mathcal{W}_{3,4}$, $\mathcal{W}\big[ \widehat{B}_{2} \big]$ and $\mathcal{W}\big[ \widehat{G}_{2} \big]$ minimal models are calculated, using the free-field approach. In Appendix \ref{sec: semi W}, we discuss the generalization of the background-charged free-field approach to the principal QDS $\mathcal{W}$ algebras related to semi-simple Lie algebras. Each simple part of a semi-simple QDS $\mathcal{W}$ algebra \textbf{decouples}, and hence the chiral characters, modular properties, fusion rules, and correlation functions of the semi-simple $\mathcal{W}$ theory are multiplications of those of the related simple $\mathcal{W}$ theories. In Appendix \ref{sec: fre cos}, we apply a coset Fock space resolution \cite{Bouwknegt:1990fb, Bouwknegt:1990wa} to express the diagonal ADE coset $\mathcal{W}$ minimal Ishibashi states \cite{Goddard:1984vk, Goddard:1986ee, Mathieu:1990dy, Bouwknegt:1992wg, Arakawa:2018iyk}. In Appendix \ref{sec: N=1}, we discuss the application of our method to the disk two-point functions of the $\mathcal{N}=1$ super-Viraosor minimal model $\mathfrak{SVir}(5,3)$. Note that for the $\mathcal{W}$ minimal models with more complicated supersymmetric chiral algebras, the resolution conjectures remained unknown due to the complexity of the embedding structure of the completely degenerate modules of the algebras.

\

\section{Preliminaries}
\label{sec: rev}

\paragraph{Concepts in simple Lie algebras and their corresponding Kac-Moody (KM) algebras.}

For a finite-dimensional Lie algebra $g$, one can always find out its maximized Abelian subalgebra $h \subset g $. The generators of $h$ is are denoted by $H^{i}$, $i=1 , \cdots ,r$, where $r$ is the rank of $g$. The rest of the $g$ generators can be determined by solving the eigenvalue equations 
\begin{equation}
    [H^{i}, E^{\alpha}] = \alpha^{i} E^{\alpha} . 
\end{equation}
The $r$ eigenvalues $\alpha^{i}$ form an $r$-dimensional vector $\alpha=(\alpha^{1},\cdots , \alpha^{r})$, which is a root vector associated to the ladder operator $E^{\alpha}$. The complete set of $g$-roots is denoted by $\Delta$ and the number of its elements is denoted by $\vert \Delta \vert$. By definition, the dimension of $g$ satisfies
\begin{equation}
    \dim(g)= r + \vert \Delta \vert . 
\end{equation} 
The roots with the first nonzero element being positive are called positive roots, whose set is denoted by $\Delta_{+}$. Note that, when $\alpha \in \Delta $, then $-\alpha \in \Delta$. Hence 
\begin{equation}
    \Delta = \Delta_{+} \cup - \Delta_{+} \quad , \quad  \vert \Delta \vert = 2 \vert \Delta_{+} \vert .
\end{equation}
For each root $\alpha$, we define its corresponding coroot as $\alpha^{\vee} = \frac{2 \alpha}{ \vert \alpha \vert^{2} }$. A simple root is defined to be a root that can't be written as the summation of two positive roots. For a Lie algebra of rank $r$, there are a total of $r$ simple roots, denoted by $\alpha_{i}$, $i=1,\cdots,r$.

\

A finite-dimension Lie algebra $g$ (whose generators are denoted by $\{ J^{a} \}$) is called simple, if it contains no proper ideal, that is, there is no proper subset of generators $\{ L^{b} \}$ such that $[J^{a}, L^{b}] \in \{ L^{b} \} $. The simple Lie algebras include four infinite series $A_{r \ge 1} \cong sl(r+1;\mathbb{C}) $, $B_{r \ge 2}= so (2r+1 ;\mathbb{C})$, $C_{r \ge 2} \cong  sp (2r ;\mathbb{C}) $, $D_{r\ge 3} \cong so(2r; \mathbb{C}) $, and five exceptional algebras $E_{6,7,8}$, $F_{4}$, and $G_{2}$.

For a simple Lie algebra $g$, there exists a unique highest root $\theta$, where the summation of linear coefficients of simple root expansions is maximized. The expansion coefficients of $\theta$ have specific notations $a_{i}$ and $a_{i}^{\vee}$, and are called the marks and the comarks respectively 
\begin{equation}
    \theta = \sum_{i=1}^{r} \: a_{i} \alpha_{i}  = \sum_{i=1}^{r} \: a_{i}^{\vee} \alpha_{i}^{\vee}. 
\end{equation}
We define a parameter $r^{\vee}$ as 
\begin{equation}
    r^{\vee} := \text{max} \big\{ a_{i} / a_{i}^{\vee} \big\} . 
\end{equation}
If $r^{\vee}=1$, the simple Lie algebra is called simply-laced, where all its roots have the same norm and it is normalized to $\vert \alpha \vert^{2}=2$. The $A_{r\ge 1}$, $D_{r\ge 3}$, and $E_{6,7,8}$ algebras are simply-laced. The $B_{r\ge 2}$, $C_{r \ge 2}$ series, and the $F_{4}$ algebra have $r^{\vee}=2$. The $G_{2}$ algebra has $r^{\vee}=3$. A finite Lie algebra $g$ is called semi-simple if it is a finite direct sum of simple Lie algebras $g= g_{1} \oplus g_{2} \oplus \cdots \oplus g_{n}$.

The weight vectors of $g$ are eigenvectors of the Abelian generators in arbitrary representations. We express the $g$ weight vectors $\lambda$ using the Dynkin labels $\lambda_{i}$
\begin{equation}
    \lambda = \sum_{i=1}^{r} \lambda_{i} \omega_{i}  =[\lambda_{1} , \lambda_{2} ,\cdots , \lambda_{r} ] , \quad (\omega_{i} , \alpha_{j}^{\vee} )= \delta_{ij} , 
\end{equation}
where $\omega_{i}$ are the fundamental weights of $g$. We define the Weyl vector $\rho$ and the dual Weyl vector $\rho^{\vee}$ of $g$ by $(\rho, \alpha_{i}^{\vee})=1$ and $(\rho^{\vee}, \alpha_{i})=1$, $\forall i$. Express them in terms of Dynkin labels 
\begin{equation}
    \rho= [1,\cdots ,1 ] , \quad \rho^{\vee} = \Big[ \frac{a_{1}}{a_{1}^{\vee}} , \cdots , \frac{a_{r}}{a_{r}^{\vee}}  \Big] .
\end{equation}
A weight vector $\lambda$ is called dominant if $\lambda_{i} \in  \mathbb{N} $, with at least one positive integer Dynkin label. The set of dominant $g$ weight vectors is denoted by $P_{+}$. Analoglously, a codominant weight vector can be defined using relation $( \lambda,\alpha_{i} ) \in \mathbb{N} $, with at least one positive integer inner product. The set of codominant weight vectors is denoted by $P_{+}^{\vee}$.

The weight lattice $P$, root lattice $Q$, and coroot lattice $Q^{\vee}$ of $g$ are defined as 
\begin{equation}
    P=  \sum_{i=1}^{r}  \mathbb{Z}\omega_{i} , \quad  Q= \sum_{i=1}^{r}  \mathbb{Z}\alpha_{i} , \quad Q^{\vee} =  \sum_{i=1}^{r}  \mathbb{Z}\alpha_{i}^{\vee}. 
\end{equation}
The number of elements in their cosets can be calculated as $\vert P/Q^{\vee} \vert = \det ( \alpha_{i}^{\vee} , \alpha_{j}^{\vee} ) $ and $\vert P/Q \vert = \det ( A_{ij} )$.

The Weyl group $W(g)$ of a finite-dimensional, bosonic Lie algebra $g$ is a finite-dimensional group, generated by $r$ simple Weyl reflections $w_{i}$
\begin{equation}
    w_{i} \lambda : = \lambda - \lambda_{i} \alpha_{i}. 
\end{equation}
For each element $w \in W(g) $, we can define its length $l(w)$ as the minimum possible number of simple Weyl reflections in its decompositions. There is a unique longest element $w_{l} \in W (g)$, and the charge conjugation of a $g$-weight vector $\lambda$ is defined as $U \lambda : =- w_{l} \lambda $. For $B_{r}$ series, $C_{r}$ series, $D_{r}$ even series, $E_{7,8}$, $F_{4}$, and $G_{2}$ the charge conjugation acts trivially on weight vectors $ U= - w_{l}=I$. For the $A_{r}$ series, $U$ reverse the order of all Dynkin labels
\begin{equation}
    U \lambda = [\lambda_{r} ,  \lambda_{r-1}  , \cdots , \lambda_{1} ] . 
\end{equation}
For the $D_{r}$ odd series, $U$ exchanges the final two Dynkin labels $\lambda_{r-1} \leftrightarrow \lambda_{r} $. For $E_{6}$, $U$ exchanges two pairs of Dynkin labels $\lambda_{1} \leftrightarrow \lambda_{5}$ and $\lambda_{2} \leftrightarrow \lambda_{4}$. We define two types of shifted Weyl actions as 
\begin{equation}
    w \ast \lambda :=  w ( \lambda + \rho ) -\rho , \quad   w \ast^{\vee} \lambda := w( \lambda+ \rho^{\vee} ) - \rho^{\vee} .
\end{equation}

\

A Kac-Moody algebra $\widehat{g}$ related to a finite Lie algebra $g$ is defined by assigning $z^{n}$, $z\in \mathbb{C}$ terms to the $g$ generators   
\begin{equation}
    H^{i}_{n}:=H^{i} \otimes z^{n} , \quad E^{\alpha}_{n} := E^{\alpha}\otimes z^{n}  , \quad z \in \mathbb{C} .
\end{equation}
The commutation relations are determined by the $g$ commutation relations, the product of $z$ powers, and the additional central extension terms. When $g$ is simple, the central extension $k$ is unique, and is called the level of the algebra $\widehat{g}(k)$ \cite{Kac:1984mq}.

For a simple KM algebra $\widehat{g}(k)$, we define its Coxeter number $h$ and the dual Coxeter numbers $h^{\vee}$  as 
\begin{equation}
   h=1+  \sum_{i=1}^{r} a_{i}  ,\quad  h^{\vee} = 1+  \sum_{i=1}^{r} a^{\vee}_{i} .
\end{equation}
A simple KM algebra has a unique central extension called the level $k$. The weight-vectors of simple KM algebras $\widehat{\lambda}$ are $(r+1)$-dimensional, where the extra Dynkin label is the zeroth Dynkin label $\lambda_{0}$, determined by 
\begin{equation*}
    \widehat{\lambda}= [\lambda_{0}, \cdots ,\lambda_{r} ] , \quad \widehat{\lambda}_{0} = k- \sum_{i=1}^{r} a^{\vee}_{i}\lambda_{i} . 
\end{equation*}
Dominant and codominant weight vectors are defined analogously to the finite cases, with the addition of an extra zeroth dimension. The set of level-$k$ dominant and codominant $\widehat{g}(k)$ weight vectors are denoted by $\widehat{P}_{+}^{k}$ and $\widehat{P}_{+}^{\vee k}$ respectively.

To study the Kac-table of the $\mathcal{W}$ minimal models, it is necessary to introduce the admissible $\widehat{g}(k)$ weight vectors. A systematic definition of the principal admissible weight vectors is given in \cite{Frenkel:1992ju}. In this work, we are only interested in the admissible $\widehat{g}(k)$ weights that take the form of \cite{Bouwknegt:1992wg}
\begin{equation}
    \widehat{\Lambda}= \alpha_{+} \widehat{\Lambda}^{(+)} + \alpha_{-} \widehat{\Lambda}^{(-)}  , \quad \widehat{\Lambda}^{(+)}\in \widehat{P}_{+} , \quad   \widehat{\Lambda}^{(-)}\in \widehat{P}_{+}^{\vee} . \label{eq: adm mod}
\end{equation}
We call their finite part admissible $g$ weight vectors, denoted by $\Lambda = (\Lambda^{(+)} ,\Lambda^{(-)} ) = \alpha_{+} \Lambda^{(+)} + \alpha_{-} \Lambda^{(-)} $. The parameters $\alpha_{\pm}$ are defined by
\begin{equation}
    \alpha_{+}^{2} = \frac{1}{k+h^{\vee}} = \frac{p'}{p} , \quad \gcd(p,p')=1, \quad  \gcd (p',r^{\vee}) =1, \quad p \ge h^{\vee} , \quad p' \ge h
     \label{eq: adm par val} .
\end{equation}

The Weyl groups of simple KM algebras $\widehat{g}(k)$ are infinite-dimensional. It is proven that the KM Weyl groups $\widehat{W}(\widehat{g})$ are semi-direct products of the finite Weyl groups $W(g)$ and translation groups $T$ along the $g$ coroot lattice
\begin{equation}
    \widehat{W}(\widehat{g}) = W(g) \ltimes T .
\end{equation}
The length for the KM Weyl group elements $l(\widehat{w})$ is defined with the involvement of the zeroth simple Weyl reflection $\widehat{w}_{0}$.

\ 

\paragraph{The free-field resolution conjectures of dominant and admissible $\widehat{g}(k)$ highest-weight modules.}

The structure of the free-field Fock space modules, and the fusion rules of the free-field theories differ significantly from the original $\widehat{g}(k)$ theories. To match the free-field properties to the original theory, it is necessary to introduce free-field chiral algebra intertwiners. An intertwiner $d$ for a Lie algebra is defined to be an object that commutes with all generators of the Lie algebra. Recall that every chiral algebra has a Virasoro subalgebra $\mathfrak{Vir} \subset \mathcal{A}$. The commutation relation $[ d,  L_{0}]=0$ indicates that the intertwiners must have conformal weight $h=0$. Further, to match the fusion rules of the free-field theory (the neutrality condition) with the original fusion rules, the intertwiners must change shift the $g$-weight vectors. Operators that shift the $g$-weight vectors with $h=0$ don't exist in local forms, but they take the form of multiple closed contour integrals of some local operators $s(z)$
\begin{equation}
    d = \oint dz_{1} \:  s_{1}(z_{1}) \cdots \oint dz_{n} \:  s_{1}(z_{n})   ,
\end{equation}
with multiple choices of the contour available \cite{Felder:1988zp}. The explicit contour is determined in the next section. These local operators $s_{i}(z_{i})$ are called the screening operators, with unity conformal weight $h=1$. From the definition of the intertwiners, the operator product expansions (OPEs) between the screening operators $s_{i}(z_{i})$ and the chiral algebra generators $W^{i}(z)$ have to be either regular or a singular total derivative term.

A common free-field approach to the simple KM $\widehat{g}(k)$ algebras is the Wakimoto-type free-field approach \cite{Wakimoto:1986gf, Bouwknegt:1990aa, Bouwknegt:1990wa}. To begin with, we rewrite the Virasoro central charge of the simple KM $\widehat{g}(k)$ using the Freudenthal-de Vries strange formula $h^{\vee} \dim(g)= 12 \vert \rho \vert^{2}$ 
\begin{equation}
    c\big[\widehat{g}(k) \big] = \frac{k \dim(g) }{ k+ h^{\vee}} = r+ \vert \Delta \vert -\frac{ 12 \vert \rho \vert^{2}  }{ k+ h^{\vee} }  . \label{eq: Vir gk}
\end{equation}
The form hinted that the simple KM $\widehat{g}(k)$ can be realized by $\vert \Delta_{+}\vert$ $\beta^{\alpha} \gamma^{\alpha}$, $\alpha \in \Delta_{+}$, bosonic ghost fields. The $r - \frac{ 12 \vert \rho \vert^{2}  }{ k+ h^{\vee} }$ part is realized by the $r$ background charged bosonic scalar fields $\phi^{i}$.

The explicit realizations of the $\widehat{g}(k)$ are complicated and model-dependent. However, the energy-stress tensor and the Cartan currents have simple and model-independent expressions \cite{Wakimoto:1986gf, Bouwknegt:1990wa} 
\begin{equation}
    T(z) = -\frac{1}{2} : \partial \phi \cdot \partial \phi : (z) - \alpha_{+} \rho \cdot i \partial^{2} \phi (z) - \sum_{\alpha\in \Delta_{+}} : \beta^{\alpha}  \partial \gamma^{\alpha}  :  (z) , \quad 
\end{equation}
\begin{equation}
    h^{i}(z) = \sum_{\alpha \in  \Delta_{+}} ( \alpha, \alpha_{i}^{\vee})  : \gamma^{\alpha} \beta^{\alpha}  : (z)  + \alpha_{+}^{-1} \alpha_{i}^{\vee} \cdot i \partial \phi^{i} .
\end{equation}
The screening operators of the WZW models take the form of \cite{Bouwknegt:1990wa}
\begin{equation}
    \widetilde{s}_{i}^{+} (z)  = \rho (e_{i}) \: : e^{-i \alpha_{+} \alpha_{i} \cdot \phi } : (z) ,
\end{equation}
where $\rho (e_{i})$ is some model-dependent polynomials of the $\beta^{\alpha}\gamma^{\alpha}$ fields \cite{Bouwknegt:1990wa}. The KM intertwiners are  $\widetilde{d}$ can be expressed as 
\begin{equation}
    \widetilde{d} =    \oint dz_{1} \: \widetilde{s}_{i_{1}}^{+} (z_{1}) \: \cdots  \:\oint dz_{n} \: \widetilde{s}_{i_{n}}^{+} (z_{n}) .
\end{equation}

\

We present a free-field resolution conjecture for the dominant and admissible simple KM modules $\widetilde{L}_{\widehat{\Lambda}}$ using the $\phi^{i} \beta^{\alpha} \gamma^{\alpha}$ Fock space modules \cite{Bouwknegt:1990wa, Bouwknegt:1991gf}. Consider a simple KM algebra at positive integer level $k\in \mathbb{Z}_{+}$, where the chiral modules are the irreducible dominant modules $\widetilde{L}_{\widehat{\Lambda}}$, $\widehat{\Lambda} \in \widehat{P}^{k}_{+}$. For dominant module $\widetilde{L}_{\widehat{\Lambda}}$, we consider the following complex $ \widetilde{C}_{ \widehat{\Lambda} }= (\widetilde{F}_{ \widehat{\Lambda} } , \widetilde{d})$
\begin{equation}
     \quad \cdots \longrightarrow \widetilde{F}_{ \widehat{\Lambda} }^{(-2)} \overset{\widetilde{d}^{-1}}{\longrightarrow}  \widetilde{F}_{ \widehat{\Lambda} }^{(-1)} \overset{\widetilde{d}^0}{\longrightarrow}  \widetilde{F}_{ \widehat{\Lambda} }^{(0)}  \overset{\widetilde{d}^1}{\longrightarrow}\widetilde{F}_{ \widehat{\Lambda} }^{(1)}  \longrightarrow \cdots ,  \label{eq: gk dom comp}
\end{equation}
where $\widetilde{F}_{\widehat{\Lambda} }^{(n)}$ is 
\begin{equation}
     \widetilde{F}_{\widehat{\Lambda} }^{(n)} = \oplus_{ \widehat{w}}^{ l(\widehat{w})=n}  \Big[ F^{\phi^{i} }_{  \widehat{w} \ast \widehat{\Lambda} } \otimes F^{\beta^{\alpha} \gamma^{\alpha}} \Big] .
\end{equation}
$\widetilde{d}$ are the $\widehat{g}(k)$ intertwiners that shift the $\widehat{g}(k)$ weight by the required amount. It is conjectured that the zeroth cohomology space of the complex $ \widetilde{C}_{ \widehat{\Lambda} }$ is isomorphic to the dominant module $\widetilde{L}_{ \widehat{\Lambda}}$ \cite{Bouwknegt:1989xa, Bouwknegt:1989jf, Bouwknegt:1990wa}
\begin{equation}
    H^{0} \big[ \widetilde{C}_{ \widehat{\Lambda} } \big]  \quad \cong \quad \widetilde{L}_{ \widehat{\Lambda}}.
\end{equation}
For the $\widehat{A}_{1}$ cases, the above conjecture is proven \cite{Bernard:1989iy}. The dominant resolution conjecture is extended to the admissible modules $\widetilde{L}_{ (\widehat{\Lambda}^{(+)}, \widehat{\Lambda}^{(-)})}$ \cite{Bouwknegt:1991gf}
\begin{equation*}
    \widehat{\Lambda} = \alpha_{+}  \widehat{\Lambda}^{(+)} + \alpha_{-}  \widehat{\Lambda}^{(-)} , \quad \widehat{\Lambda}^{(+)} \in  \widehat{P}_{+}^{p-h^{\vee}} ,\quad  \widehat{\Lambda}^{(-)} \in \widehat{P}_{+}^{ \vee\: p'-h} , \label{eq: adm weight}
\end{equation*}
\begin{equation}
    \alpha_{+}^{2} = \frac{1}{k+h^{\vee}} = \frac{p'}{p} , \quad \gcd(p,p')=1, \quad  \gcd (p',r^{\vee}) =1, \quad p \ge h^{\vee} , \quad p' \ge h
     \label{eq: adm par val} .
\end{equation}
The resolution complexes $\widetilde{C}_{(\widehat{\Lambda}^{(+)},\widehat{\Lambda}^{(-)})}$ for the admissible modules are obtained by replacing the Fock space modules in the dominant resolution complexes by 
\begin{equation}
    \widetilde{F}^{(n)}_{(\widehat{\Lambda}^{(+)},\widehat{\Lambda}^{(-)})} =  \oplus_{ \widehat{w}}^{ l(\widehat{w})=n}  \Big[ F^{\phi^{i} }_{  ( \widehat{w} \ast \widehat{\Lambda}^{(+)},  \widehat{\Lambda}^{(-)} ) } \otimes F^{\beta^{\alpha} \gamma^{\alpha}} \Big] , 
\end{equation}
and the admissible module $\widetilde{L}_{(\widehat{\Lambda}^{(+)},\widehat{\Lambda}^{(-)})}$ is conjectured to be isomorphic to the zeroth cohomology space of the complex $\widetilde{C}_{(\widehat{\Lambda}^{(+)},\widehat{\Lambda}^{(-)})}$. 

\

\paragraph{Background-charged free boson approach to principal QDS $\mathcal{W} $ algebras.}

We begin by defining the quantum Drinfeld-Sokolov (QDS) $\mathcal{W} $ algebras, related to a simple KM algebra $\widehat{g}$. First, one begins by selecting a subalgebra of the KM algebra $\widehat{g}' \subset \widehat{g}$. Next, one introduce an one-dimensional representation of $\widehat{g}'$, and impose the constraint condition $ x \sim \chi (x)$, $\forall x \in \widehat{g}'$ \cite{Bershadsky:1989mf, Feigin:1990pn, Bouwknegt:1992wg, Kac:2003jh}. The triple $( \widehat{g} , \widehat{g} ' , \chi )$ defines a $\mathcal{W}$ as the zeroth BRST cohomology of the triple
\begin{equation}
    \mathcal{W} \big[  ( \widehat{g} , \widehat{g} ' , \chi ) \big] \quad \cong \quad H_{Q}^{(0)} ( \widehat{g} , \widehat{g} ' , \chi ). 
\end{equation}
The $\mathcal{W}$ algebras considered in this work are obtained from imposing the principal $A_{1}$ embedding condition
\begin{equation}
    \chi\big[  e_{\alpha} (z) \big]   = \begin{cases}
        1 &  \forall \alpha_{i}  \\  0 & \text{otherwise} .
    \end{cases}
\end{equation}
where $e_{\alpha}(z)$ are the Chevalley currents. The Virasoro central charge of the principle $A_{1}$ embedding $\mathcal{W} \big[ \widehat{g}(k)  \big] $ is 
\begin{equation}
    c = r- 12 \vert \alpha_{+} \rho + \alpha_{-} \rho^{\vee} \vert^{2} .  
\end{equation}
It is proven that the conformal weights of the $W^{i}(z)$ chiral generators given by $h^{i}=e_{i}+1$, $i=1,\cdots, r$, where $e_{i}$ are the exponents of the Lie algebra $g$. In other words, the $W^{i}(z)$ conformal weights $h^{i}$ are the orders of the independent Casimir of $g$. 

\ 

The focus of this work is on the free-field approaches to the QDS $\mathcal{W}$ minimal models. Recall that a free-field approach to the simple KM algebra $\widehat{g}(k)$ is the Wakimoto approach using $r$ background-charged fields $\phi^{i}$, and $\vert \Delta_{+} \vert$ $\beta^{\alpha} \gamma^{\alpha}$, $\alpha\in \Delta_{+}$ bosonic ghost fields. Next, we introduce the following lemma  \cite{Bershadsky:1989mf}
\begin{equation}
     H_{Q}^{(0)} \big(  F_{\beta \gamma} \otimes F_{\text{bc}} \big)  \cong  \mathbb{C}. 
\end{equation}
This lemma indicates that the $r$ background-charged fields $\phi^{i}$ provides a natural free-field approach for the theories with the QDS $\mathcal{W}$ algebra. The free-field $\mathcal{W}\big[ \widehat{g} (k) \big] $ energy-stress tensor is given by \cite{Bouwknegt:1992wg}
\begin{equation}
    T(z) = -\frac{1}{2} :  \partial \phi \cdot \partial \phi : (z) - ( \alpha_{+} \rho + \alpha_{-} \rho^{\vee} ) \cdot i \partial^{2} \phi (z) .  \label{eq: W fre T}
\end{equation}
with complicated realizations of other $W^{i}(z)$ chiral generators. The $\mathcal{W}\big[ \widehat{g} (k) \big] $ screening operators are 
\begin{equation}
    s^{\pm}_{i} (z) = \: :e^{-i \alpha_{\pm} \alpha_{i} \cdot \phi } : (z).
\end{equation}
with the $\mathcal{W}\big[ \widehat{g} (k) \big] $ intertwiners are denoted by $d$.

Let $ C_{( \Lambda^{(+)} , \Lambda^{(-)} )}= ( F_{( \Lambda^{(+)} , \Lambda^{(-)} )} ,d)$ be the BRST image of the $\widehat{g}(k)$ admissible resolution complex $\widetilde{C}_{( \widehat{\Lambda}^{(+)} ,  \widehat{\Lambda}^{(-)} )}$, where $d$ are $\mathcal{W} \big[ \widehat{g} \big] (p,p')$ intertwiners
\begin{equation}
    C_{( \Lambda^{(+)} , \Lambda^{(-)} )} : \quad  \quad \cdots \longrightarrow F_{ \widehat{\Lambda} }^{(-2)} \overset{d^{-1}}{\longrightarrow}  F_{ \widehat{\Lambda} }^{(-1)} \overset{d^0}{\longrightarrow}  F_{ \widehat{\Lambda} }^{(0)}  \overset{d^1}{\longrightarrow} F_{ \widehat{\Lambda} }^{(1)}  \longrightarrow \cdots ,
\end{equation}
where $F_{ \widehat{\Lambda} }^{(i)}$ is
\begin{equation}
    F_{ \widehat{\Lambda} }^{(i)} = \bigoplus_{\widehat{w}}^{l(\widehat{w})=i} F_{\widehat{w} \ast \widehat{\Lambda} }^{\phi} . 
\end{equation}
If the KM highest weight vector $\widehat{\Lambda}$ is dominant or admissible vector taking the form of (\ref{eq: adm weight}), the zeroth cohomology space of the BRST image complex $(F_{\widehat{\Lambda}} , d )$ is either zero or irreducible modules $L_{( \Lambda^{(+)} , \Lambda^{(-)} ) }$ \cite{Bouwknegt:1991gf, Frenkel:1992ju}
\begin{equation}
    L_{( \Lambda^{(+)} , \Lambda^{(-)} )} \quad \cong \quad H^{0} \big[ C_{( \Lambda^{(+)} , \Lambda^{(-)} )} \big].
\end{equation}

\ 

\paragraph{Rational $\mathcal{W}\big[ \widehat{g} \big](p,p')$ minimal models.}

When the admissible KM highest weight vector $\widehat{\Lambda}$ is parameterized by (\ref{eq: adm par val}), the corresponding QDS $\mathcal{W}$ algebra are denoted by $\mathcal{W}\big[ \widehat{g} \big](p,p')$. We consider rational $\mathcal{W}\big[ \widehat{g} \big](p,p')$ minimal models. The Kac-tables of the rational $\mathcal{W}\big[ \widehat{g} \big](p,p')$ minimal models are \cite{Bouwknegt:1992wg} 
\begin{equation}
    \Big\{  L_{( \Lambda^{(+)} , \Lambda^{(-)})} \quad \vert \quad \widehat{\Lambda}^{(+)}\in  \widehat{P}^{p-h^{\vee}}_{+} ,\: \widehat{\Lambda}^{(-)}\in  \widehat{P}^{ \vee p'-h}_{+}  \Big\} , \label{eq: W min Kac tab}
\end{equation}
with a Weyl symmetry identification
\begin{equation}
    L_{( \Lambda^{(+)} , \Lambda^{(-)})}  \: \cong  \:  L_{( \widehat{w} \ast \widehat{\Lambda}^{(+)} + p \beta , \:  \widehat{w} \ast^{\vee} \widehat{\Lambda}^{(-)} + p' \beta )} \: , \quad \beta \in Q^{\vee}_{L} , \label{eq: Weyl sym}
\end{equation}
where $Q^{\vee}_{L}$ is the long root lattice of $g$. The primary conformal weight $h_{(\Lambda^{(+)} , \Lambda^{(-)})}$ of the irreducible module $L_{(\Lambda^{(+)} , \Lambda^{(-)})}$ is given by 
\begin{equation}
    h_{(\Lambda^{(+)} , \Lambda^{(-)})} = \frac{1}{2} ( \Lambda , \Lambda  +2 \alpha_{+} \rho + 2\alpha_{-} \rho^{\vee} ) .  
\end{equation}

The $\mathcal{W}\big[ \widehat{g} \big](p,p')$ modular $S$ matrices take the form of \cite{Lukyanov:1990tf, Mathieu:1990dy, Frenkel:1992ju, Arakawa:2016lef}
\begin{equation*}
    S_{( \Lambda^{(+)}, \Lambda^{(-)} ) (\mu^{(+)} , \mu^{(-)})}  =  \frac{1}{\sqrt{(pp')^r }} \Big\vert  \frac{P}{Q^{\vee}} \Big\vert^{-\frac{1}{2}}   e^{2\pi i ( \Lambda^{(+)} \mu^{(-)} +\Lambda^{(-)} \mu^{(+)} ) } 
\end{equation*}
\begin{equation}
   \sum_{w\in W(g)} \epsilon(w) e^{- \frac{2\pi i p'}{p} \big[ (\Lambda^{(+)}+ \rho ),\: w(\mu^{(+)} + \rho )  \big]  } \sum_{w' \in W(g) } \epsilon(w') e^{- \frac{2\pi i p}{p'} \big[ (\Lambda^{(-)}+ \rho ),\: w'(\mu^{(-)} + \rho )  \big]  } . \label{eq: mod S conj}
\end{equation}
The fusion rules of RCFTs can be obtained from the Verlinde formula \cite{Dijkgraaf:1988tf, Verlinde:1988sn}
\begin{equation}
    N_{ij}^{k} =  \sum_{l} \frac{ S_{il} S_{jl} S_{kl}^{-1}  }{ S_{0l}} . \label{eq: Ver for}
\end{equation}
The fusion rules related to the identity module are called obvious rules in this work.

\ 

\paragraph{Cardy boundary states.}

The disk correlation functions considered are disk bulk two-point functions, with Cardy boundary conditions. The concept of Cardy boundary states are reviewed in this paragraph \cite{Cardy:1989ir}.

Consider a CFT$_2$ defined on the upper-half plane (UHP), the chiral generators satisfy the gluing conditions on the boundary (only the chiral symmetry preserved cases are considered) \cite{Cardy:1984bb, Recknagel:1997sb, Recknagel:1998ih} 
\begin{equation}
    T(z)= \widebar{T}(\widebar{z}) , \quad W^{i}(z)=  \Omega\widebar{W}^{i}(\widebar{z}) ,
\end{equation}
where $\Omega$ is an outer automorphism of the chiral algebra $\mathcal{A}$. Mapping the UHP to the unit disk, and assigning a boundary state $\vert b \rangle_{\Omega}$ to its boundary, we obtain the conditions
\begin{equation}
    (L_{n}- \widebar{L}_{-n}) \vert b \rangle_{\Omega} = \big[ W_{n}^{i}- (-1)^{h_{i}} \Omega \widebar{W}^{i}_{-n} \big] \vert b \rangle_{\Omega}  =0 . \label{eq: Ishi con}
\end{equation}
For each $\mathcal{A}$ representation $L_{i}$ (not necessarily irreducible), the solutions to the conditions (\ref{eq: Ishi con}) are the $\Omega$-twisted Ishibashi states $\vert  L_{i} \rangle \rangle_{\Omega} $ \cite{Ishibashi:1988kg, Onogi:1988qk} 
\begin{equation}
    \vert L_{i} \rangle \rangle_{\Omega} = \sum_{N=0}^{\infty} \vert i ,N \rangle \otimes  V_{\Omega} U \vert i ,N \rangle ,
\end{equation}
where $\vert i,N \rangle$ denotes the orthonormal basis of $L_{i}$. $U$ is an anti-unitary isomorphism mapping a representation to its charge conjugation. $V_{\Omega}$ is a unitary isomorphism between the chiral modules induced from $\Omega$ \cite{Recknagel:1997sb, Recknagel:2013uja}.

The Ishibashi states are not the physical boundary states yet, since physical boundary states have to satisfy further physical constraint conditions \cite{Cardy:1989ir}. The constraint condition originates from equivalence methods of calculating the annulus partition functions, and the form of the open-sector Hilbert space between two boundaries of RCFTs \cite{Cardy:1989ir}
\begin{equation}
    \mathcal{H}^{b_{1}b_{2}} = \bigoplus_{i }\: n_{ b_{1}b_{2} }^{i} L_{i} , \quad  n_{ b_{1}b_{2} }^{i}\in \mathbb{N}.
\end{equation}
In this work, we don't discuss the solutions to the constraint conditions in the RCFTs, but we simply pick the simplest known solution, which is the celebrated Cardy boundary conditions \cite{Cardy:1989ir}. The expressions are given by 
\begin{equation}
    \vert b_{i} \rangle_{\Omega} =  \sum_{j} \frac{S_{ij} }{\sqrt{S_{0j}}}  \vert  L_{j}  \rangle \rangle_{\Omega} .
\end{equation}

\ 

\section{Universal results}
\label{sec: Uni re}

\paragraph{Free-field expressions of the $\mathcal{W}$ minimal Ishibashi states.}

Consider diagonal or charge conjugated simple $\mathcal{W}\big[ \widehat{g}\big](p,p')$ minimal models. We apply the free-field resolution conjecture to Ishibashi states
\begin{equation}
    \vert  L_{( \Lambda^{(+)} ,\Lambda^{(-)} )} \rangle \rangle_{\Omega}  = \sum_{  \widehat{w} \in \widehat{W} }  \theta_{\widehat{w}}  \vert  F_{ ( \widehat{w} \ast \widehat{\Lambda}^{(+)} ,  \widehat{\Lambda}^{(-)} )  }  \rangle \rangle_{\Omega'}   ,  \quad \Omega=U,I ,  \label{eq: Ishi free} 
\end{equation}  
where $\theta_{\widehat{w}} \in U(1)$ are undetermined phases. We write the dual Ishibashi states as 
\begin{equation}
    _{\Omega} \langle \langle  L_{( \Lambda^{(+)} ,\Lambda^{(-)} )} \vert = \sum_{  \widehat{w} \in \widehat{W} }  \theta_{\widehat{w}}' \:  _{\Omega'} \langle \langle F_{ ( \widehat{w} \ast \widehat{\Lambda}^{(+)} ,  \widehat{\Lambda}^{(-)} )  }  \vert . 
\end{equation}
The undetermined phases in the two equations satisfy $\theta_{\widehat{w}} \theta_{\widehat{w}}' =(-1)^{l(\widehat{w})}$, to reproduce the correct chiral character from the overlaps of the free-field expressions. The form of $\vert  F_{ ( \widehat{w} \ast \widehat{\Lambda}^{(+)} ,  \widehat{\Lambda}^{(-)} )  }  \rangle \rangle_{\Omega'}$ is 
\begin{equation}
    \vert F_{ (  \widehat{w} \ast \widehat{\Lambda}^{(+)} , \widehat{\Lambda}^{(-)}) } \rangle \rangle_{\Omega'} = \exp{ \Big( \frac{-1}{n} a_{-n}^{i} \cdot V_{\Omega'} \cdot \widebar{a}_{-n}^{i}  \Big) }  \Big\vert  p \otimes \widebar{p}  \Big\rangle. 
\end{equation}
$p$ is the finite part of the $\alpha_{+} \widehat{w} \ast \widehat{\Lambda}^{(+)} + \alpha_{-} \widehat{\Lambda}^{(-)}$ and $\widebar{p}$ is determined by imposing the Virasoro gluing condition. The free-field gluing automorphism $V_{\Omega}'$ is determined by imposing the $\mathcal{W}$ gluing conditions to the free-field Ishibashi states. Generically speaking, it is necessary to know the explicit forms of the free-field $\mathcal{W}$ chiral generators and the action of the $V_{\Omega}$ of the $\mathcal{W}$ chiral generators. For the diagonal and charge conjugated models considered in this work, it is not necessary to know the explicit free-field $\mathcal{W}$ realizations. Note that, vectors in different Ishibashi states belong to different Fock space modules, and the action of the chiral generators does not change the Fock space modules. Hence
\begin{equation}
    (W^{i}_{n} - \Omega (-1)^{h_{i}} \widebar{W}_{-n}^{i} ) \vert L_{(\Lambda^{(+)} , \Lambda^{(-)} )} \rangle \rangle_{\Omega} =0  \quad \Longrightarrow \quad  (W^{i}_{n} - \Omega (-1)^{h_{i}} \widebar{W}_{-n}^{i} ) \vert F_{( \widehat{w} \ast \widehat{\Lambda}^{(+)} , \widehat{\Lambda}^{(-)} )} \rangle \rangle_{\Omega'}  =0  , 
\end{equation}
for all $\widehat{w} \in \widehat{W} $.

We begin with imposing the simplest Virasoro gluing conditions to the free-field Ishibashi states. Recall the form of free-field $\mathcal{W} \big[ \widehat{g}(k) \big]$ energy-stress tensor (\ref{eq: W fre T}), imposing the Virasoro gluing conditions leads to conditions similar to those of \cite{Kawai:2002vd, Kawai:2002pz} 
\begin{equation*}
    V_{\Omega'}^{2}-I =0 ,  \quad ( \alpha_{+} \rho + \alpha_{-} \rho^{\vee} )   \cdot (V_{\Omega'}-I) =0 ,  
\end{equation*}
\begin{equation}
    ( p +  \alpha_{+} \rho + \alpha_{-} \rho^{\vee}  ) \cdot V_{\Omega'}  + \widebar{p} + ( \alpha_{+} \rho + \alpha_{-} \rho^{\vee} ) =0  . \label{eq: Vir glu con}
\end{equation}
Next, we consider imposing the $\mathcal{W}$ gluing conditions. The $W^{i}(z)$ chiral generators other than the energy-stress tensor $T(z)$ are Virasoro primary operators (this is ensured by the $W^{i}$ generator projection theorem). Hence, the following commutation relations hold
\begin{equation}
    [L_{m}, W^{i}_{n}] = \big[ (h^{i}-1)m - n  \big]   W_{m+n}^{i} .
\end{equation}
Following the same calculation for the $\mathcal{W}_{3}$ case in \cite{Caldeira:2003zz}, we write the $\mathcal{W}$ gluing conditions as
\begin{equation}
    \frac{1}{(h^{i}-1) n } (L_{n}- \widebar{L}_{-n}) \big[ W_{0}^{i} - (-1)^{h_{i}} \Omega \widebar{W}^{i}_{0} \big] \vert L_{( \Lambda^{(+)}, \Lambda^{(-)} )} \rangle  \rangle_{\Omega}=0 . \label{eq: glu cond}
\end{equation}
(\ref{eq: glu cond}) indicates that all the $W^{i}$ gluing conditions are satisfied if the Virasoro gluing condition are satisfied, and the anti-holomorphic momentum $\widebar{p}$ provide the correct $\widebar{w}^{i}$ eigenvalue.

Next, we need to determine how the $W^{i}(z)$ eigenvalues $w^{i}$ change under the relation
\begin{equation}
    \widebar{p} = -p -2\alpha_{+} \rho -2 \alpha_{-} \rho^{\vee} .\label{eq: p-pbar}
\end{equation}
We conjecture that $w^{i}$ eigenvalues are the standard symmetric polynomial of order $i$ of the variables $\theta_{j} = ( \Lambda +\alpha_{+} \rho + \alpha_{-} \rho^{\vee} , \epsilon_{j} ) $, where $\{ \epsilon_{j} \}$ is the set of weight of vector representation of $g$. This conjecture is a direct generalization of the $w^{3}$ eigenvalue shown in \cite{Bouwknegt:1992wg}. The $w^{i}$ eigenvalues changed under the action $\Lambda \to -\Lambda -2\alpha_{+} \rho- 2\alpha_{-} \rho^{\vee}$ as $w^{i} \to (-1)^{h^{i}}  w^{i}$. This is the expected the $w^{i}$ eigenvalue shift under the charge conjugation
\begin{equation}
   U : \quad   w^{i} \to (-1)^{h^{i}}  w^{i} ,
\end{equation}
since the trivial $W^{i}_{0}$ gluing conditions is $ \big[ W^{i}_{0} - (-1)^{h^{i}} \widebar{W}_{0} \big] \vert L_{(\Lambda^{(+)} , \Lambda^{(-)} )} \rangle \rangle_{I} =0 $, with the corresponding solution being and Ishibashi state in the charge conjugated model. Hence, we can understand that the shift $\Lambda \to -\Lambda -2\alpha_{+} \rho- 2\alpha_{-} \rho^{\vee}$ is a form of charge conjugation in the free-field realization.

From the above discussion, we conclude that the trivial $W^{i}_{0}$ gluing conditions 
\begin{equation}
    \big[ W^{i}_{0}-(-1)^{h^{i}}\widebar{W}^{i}_{0} \big] \vert L_{ (\Lambda^{(+)} , \Lambda^{(-)}) } \rangle \rangle_{I} =0 ,  
\end{equation}
are satisfied by the free-field Ishibashi states with $V_{\Omega'}=I$, and $\widebar{p} = -p -2\alpha_{+} \rho -2 \alpha_{-} \rho^{\vee}  $, for all $W^{i}(z)$ generators. The charge conjugated $W^{i}_{0}$ gluing conditions 
\begin{equation}
    ( W^{i}_{0}-\widebar{W}^{i}_{0} ) \vert L_{ (\Lambda^{(+)} , \Lambda^{(-)}) } \rangle \rangle_{U} =0 , 
\end{equation}
are satisfied by the relation $V_{\Omega'}=U$, and $\widebar{p} = - U \cdot ( p + \alpha_{+} \rho + \alpha_{-} \rho^{\vee} ) -\alpha_{+} \rho - \alpha_{-} \rho^{\vee} = w_{l}\cdot ( p + \alpha_{+} \rho + \alpha_{-} \rho^{\vee} ) -\alpha_{+} \rho - \alpha_{-} \rho^{\vee}  $,  because the $p-\widebar{p}$ relation is a Weyl symmetry identification (\ref{eq: Weyl sym}), leaving the $w^{i}$ eigenvalues invariant.

\ 

\paragraph{Generalized Coulomb-gas formalism.}

The Coulomb-gas formalism computes correlation functions using free-field vertex operators and the insertion of chiral intertwiners \cite{Dotsenko:1984nm, Dotsenko:1984ad, Dotsenko:1985hi, Felder:1988zp}. For a given bulk primary operator $\Phi_{(i,\widebar{i})} (z_{i} , \widebar{z}_{\widebar{i}})$ in the original CFT$_2$, we introduce a pair of primary free-field vertex operators to represent it \cite{Dotsenko:1984nm, Felder:1988zp} 
\begin{equation}
    \Phi_{(i,\widebar{i})} (z_{i} , \widebar{z}_{\widebar{i}})  \quad \longrightarrow \quad V_{i}(z_{i})\widebar{V}_{\widebar{i}}(\widebar{z}_{\widebar{i}}),
\end{equation}
where $V_{i}(z_{i})=\: : e^{i p_{i} \phi } : (z_{i})$ and $\widebar{V}_{\widebar{i}}(\widebar{z}_{\widebar{i}})=\: : e^{i \widebar{p}_{\widebar{i}} \widebar{\phi} } : (\widebar{z}_{\widebar{i}})$ are required to process the same $\mathcal{A}$ and $\widebar{\mathcal{A}}$ eigenvalues as the original operator $\Phi_{(i,\widebar{i})} (z_{i} , \widebar{z}_{\widebar{i}})$. In the models considered in this work, the eigenvalues include the conformal weight $h$, the $w^{i}$ eigenvalues for the $\mathcal{W}$ minimal models, and the Dynkin labels for the WZW models. The choices for the free-field vertex operators of one bulk operator are not unique.

For the disk correlation functions $_{\Omega}\langle b \vert  \prod_{(i,\widebar{i})}^{n} \Phi_{(i,\widebar{i})} (z_{i},\widebar{z}_{\widebar{i}}) \vert 0 \rangle$ associated with the Cardy boundary states. Substituting the explicit form of the Cardy boundary state yields the following expansion
\begin{equation}
    _{\Omega}\langle b_{a} \vert  \prod_{(i,\widebar{i})}^{n} \Phi_{(i,\widebar{i})} (z_{i},\widebar{z}_{\widebar{i}}) \vert 0 \rangle = \sum_{j} \frac{(S_{aj}^{-1})}{\sqrt{S_{0j}}}\:  _{\Omega}\langle \langle L_{j} ,b \vert \prod_{(i,\widebar{i})}^{n} \Phi_{(i,\widebar{i})} (z_{i},\widebar{z}_{\widebar{i}}) \vert 0 \rangle . \label{eq: dis cor exp}
\end{equation}
When the explicit modular $S$ matrix is non-zero, and the fusion rules of the original CFT$_2$ are allowed, the contribution in the expansion is non-zero. To evaluate these components, we adopt the methodology introduced in \cite{Kawai:2002pz, Caldeira:2003zz, Hemming:2004dm}, where a contribution $_{\Omega}\langle\langle L_{j} ,b \vert \prod_{(i,\widebar{i})}^{n}  \Phi_{(i,\widebar{i})} (z_{i},\widebar{z}_{\widebar{i}}) \rangle$ is \textbf{assumed} to be represented by   
\begin{equation}
    _{\Omega}\langle\langle F_{p} ,b \vert  \prod_{i,\widebar{i}=1}^{n} V_{i} (z_{i})  \widebar{V}_{\widebar{i}} (\widebar{z}_{\widebar{i}}) \rangle_{D^{2} } = \delta \widebar{\delta} \prod_{i<j} z_{ij}^{p_{i} \cdot p_{j}} \prod_{\widebar{i}<\widebar{j}} \widebar{z}_{\widebar{i}\widebar{j}}^{\widebar{p}_{\widebar{i}} \cdot \widebar{p}_{\widebar{j}}} \prod_{i,\widebar{j}} (1- z_{i} \widebar{z}_{j})^{p_{i} \cdot V_{\Omega} \cdot \widebar{p}_{\widebar{j}} } . \label{eq: disk}
\end{equation}
Here, the neutrality conditions are $\delta= \delta_{\sum_{i} p_{i} -p ,0 }$ and $\widebar{\delta}= \delta_{\sum_{\widebar{i}} \widebar{p}_{\widebar{i}} -\widebar{p} ,0  }$, where $p$ and $\widebar{p}$ are the momentum carried by the Fock space Ishibashi state.

\

\paragraph{The procedure of disk two-point function calculations.}

We only consider calculations of disk two-point functions, and there are two crucial techniques in the calculations.
\begin{itemize}
    \item We send the position $z_{2} \to 0$, which simultaneously send $\widebar{z}_{2} \to \infty$ from the doubling trick on the disk. We perform an inversion coordinate transformation $\widebar{z}_{2} \to 0$ and absorb the Jacobian into the normalization. 

    \item The vertex operators and the screening operators are chosen to have radial ordering $0= \vert z_{2} \vert <  \vert z^{n} \vert <  \vert z^{n-1} \vert < \cdots < \vert  z_{1} \vert  < 1  $, with the contours shown in the Figure 1 in \cite{Felder:1988zp}. These contours are equivalent to the Pochhammer contours, with the integral is 
    \begin{equation}
        \oint_{P} dz^{1}  \cdots \oint_{P} dz^{n-1} \oint_{P} dz^{n} .
    \end{equation}
    If we deform the contour integrals to the line segment integrals on real axis as \cite{Dotsenko:1984nm, Dotsenko:1984ad}
    \begin{equation}
        \int_{0}^{z_{1}} dz^{1} \cdots \int_{0}^{z^{n-2}} dz^{n-1}  \int_{0}^{z^{n-1}} dz^{n}   . 
    \end{equation}
    However, we avoid the deformation to line segments since the endpoint divergence could exist. 
\end{itemize}

\ 

We have determined the form of disk correlation function integrals, the integrals can be calculated analytically using the following information.

On $\mathbb{C}^{n}$, a Lauricella hypergeometric function $F_{D}^{(n)} \Big[ a, b_{1}, \cdots , b_{n}; c ; z_{1} , \cdots , z_{n} \Big] $, $z_{i} \in \mathbb{C}$ is defined as 
\begin{equation}
    F_{D}^{(n)} \Big[ a , b_{1}, \cdots , b_{n}; c ; z_{1} , \cdots , z_{n} \Big] = \sum_{k_{i}  \in  \mathbb{N}}  \frac{ (a)_{\sum_{i=1}^{n} k_{i}  }  \prod_{i=1}^{n} (b)_{k_{i}}  }{(c)_{ \sum_{i=1}^{n} k_{i}} } \prod_{i=1}^{n} \frac{z_{i}^{k_{i}} }{k_{i}!} , \label{eq: Tay exp FD}
\end{equation}
which is convergent when $\max{ \vert z_{i} \vert  } <1 $. $(a)_{k}= \Gamma(a+k)/ \Gamma(a)$ is the Pochhammer symbol. The $F_{D}^{(n)}$ functions admit Euler-type integral expressions 
\begin{equation*}
     F_{D}^{(n)} \Big[ a , b_{1}, \cdots , b_{n}; c ; z_{1} , \cdots , z_{n} \Big] = 
\end{equation*}
\begin{equation}
    \frac{\Gamma(c)}{\Gamma (a) \Gamma(c-a)}\int_{0}^{1} dt\: t^{a-1} (1-t)^{c-a-1} \prod_{i=1}^{n} (1-z_{i} t)^{-b_{i}} ,
\end{equation}
where $\text{Re}(c-a) >0 $ and $\text{Re}(a) >0 $. The integral is divergent if the previous two conditions are not satisfied. The endpoint divergence doesn't exist for the Pochhammer contour integral expressions of the Lauricella hypergeometric functions $F_{D}^{(n)}$ (with integrand dependent factors eliminated)
\begin{equation*}
     F_{D}^{(n)} \Big[ a , b_{1}, \cdots , b_{n}; c ; z_{1} , \cdots , z_{n} \Big] = 
\end{equation*}
\begin{equation}
    \frac{\Gamma(c)}{\Gamma (a) \Gamma(c-a)}\oint_{P} dt\: t^{a-1} (1-t)^{c-a-1} \prod_{i=1}^{n} (1-z_{i} t)^{-b_{i}} ,
\end{equation}
where the $\oint_{P}$ denotes the Pochhammer contour integral around the two branching points $0$ and $1$. The above integral expression and the Taylor expansion (\ref{eq: Tay exp FD}) is valid $\forall a$, $b_{i}$, $c \in \mathbb{C}$.

With the above knowledge, disk two-point functions can be obtained by the repeated application of the integral expressions and the Taylor expansions of the Lauricella hypergeometric function $F_{D}^{(n)}$.

\ 

\paragraph{Disk one-point functions.}

The calculations of disk one-point functions don't require any insertions of the chiral intertwiners. Consider a minimal model with diagonal and charge conjugation modular invariants. Disk one-point functions in $\Omega$ permuted RCFTs, with Cardy boundary conditions are 
\begin{equation*}
    _{U\Omega}\langle b_{( \nu^{(+)} , \nu^{(-)}}) \vert \Phi_{( \Lambda^{(+)} , \Lambda^{(-)})}^{\Omega} ( z_{1} , \widebar{z}_{1} ) \vert 0 \rangle = 
\end{equation*}
\begin{equation}
    \frac{ S^{-1}_{ ( \nu^{(+)} , \nu^{(-)} ) (\Lambda^{(+)} , \Lambda^{(-)} ) } }{\sqrt{ S_{ 0(\Lambda^{(+)} , \Lambda^{(-)} ) } }} \:  _{U\Omega} \langle \langle L_{ ( \Lambda^{(+)} , \Lambda^{(-)}) } \vert \Phi_{( \Lambda^{(+)} , \Lambda^{(-)})}^{\Omega} ( z_{1} , \widebar{z}_{1} ) \vert 0 \rangle .
\end{equation}
The free-field realizations of the disk one-point functions in diagonal models are 
\begin{equation*}
    _{U}\langle\langle L_{ ( \Lambda^{(+)} , \Lambda^{(-)}) } \vert V_{ ( \Lambda^{(+)} , \Lambda^{(-)}) }  ( z_{1} ) \widebar{V}_{ ( w_{l} \ast \Lambda^{(+)} , w_{l} \ast^{\vee} \Lambda^{(-)}) } ( \widebar{z}_{1}) \vert 0 \rangle 
\end{equation*}
\begin{equation}
     =(1- \vert z_{1} \vert^{2})^{-  \Lambda  \cdot w_{l}  \cdot \big[ w_{l} \cdot (\Lambda+ \alpha_{+} \rho +\alpha_{-} \rho^{\vee} ) - (\alpha_{+} \rho +\alpha_{-} \rho^{\vee} )  \big]  } .
\end{equation}
The free-field realizations of that in charge-conjugate models are 
\begin{equation*}
    _{I}\langle\langle L_{ ( \Lambda^{(+)} , \Lambda^{(-)}) } \vert V_{ ( \Lambda^{(+)} , \Lambda^{(-)}) }  ( z_{1} )  \widebar{V}_{ ( \Lambda^{(+)} , \Lambda^{(-)}) }^{\dag} ( \widebar{z}_{1}) \vert 0 \rangle 
\end{equation*}
\begin{equation}
   = (1-\vert z_{1} \vert^{2})^{-  \Lambda  \cdot ( \Lambda + 2\alpha_{+} \rho +2\alpha_{-} \rho^{\vee})   }  = (1-\vert z_{1} \vert^{2})^{-2 h_{ ( \Lambda^{(+)} , \Lambda^{(-)})}} . 
\end{equation}

\ 

\section{Disk two-point function calculations in some $\mathcal{W} \big[ \widehat{g} \big] (p,p')$ minimal models}
\label{Sec: Disk 2pt}

In this subsection, we show detailed examples of calculations of the modular $S$ matrices, fusion rules, and the disk two-point functions with Cardy boundary states in some $\mathcal{W} \big[ \widehat{g} \big] (p,p')$ minimal models. Note that, not all choices of the free-field vertex operators provide us with physically admissible exponents, we need to verify our choice by hand. Also, in this work, we only calculate the contribution from each Ishibashi state, with the final summation omitted.

\

\subsection{Virasoro minimal models}

The minimal Virasoro algebras $\mathfrak{Vir}(p,p')$ are the principal QDS $\mathcal{W} $ algebras of the $\widehat{A}_{1}(k)$ algebras, with $k+2 = p /p'$. The Virasoro minimal Kac-table is
\begin{equation}
     \big\{  L_{(r,s)}  \: \vert \:  1 \le r\le p-1 , \quad 1 \le s\le p'-1 \big\}.
\end{equation}
The labels are related to the Dynkin labels of $\widehat{A}_{1}$ highest-weight by $r= \Lambda^{(+)}_{1} +1$ and $s= \Lambda^{(-)}_{1} +1$. The Weyl symmetry on the Kac-table is $L_{(r,s)} \equiv L_{(p-r, p'-s)}$. The $A_{1}$ Weyl group has only two elements $W=\{ I, s_{1} =w_{l} \}$. Hence, Virasoro modular $S$ matrices have a simple expression
\begin{equation}
    S_{(r,s)(r',s')} = \sqrt{ \frac{8}{pp'} } (-1)^{(1+rs'+sr')}\sin{\Big( \pi  \frac{p'}{p} rr' \Big)} \sin{\Big( \pi \frac{p}{p'} ss' \Big)} .
\end{equation}
The Virasoro minimal fusion rules can be obtained from the Kac-Walton formula 
\begin{equation}
    L_{(r_{1},s_{1})} \times L_{(r_{2},s_{2})} = \sum_{r_{3} = \vert r_{1} -r_{2} \vert +1  }^{ \min (r_{1}+r_{2}-1 , 2p-r_{1}-r_{2}-1 ) } \:   \sum_{s_{3} = \vert s_{1} -s_{2} \vert +1  }^{ \min (s_{1}+s_{2}-1 , 2p'-s_{1}-s_{2}-1 ) } L_{(r_{3},s_{3})} , 
\end{equation}
where the summation of $r_{3}$ and $s_{3}$ jumps by $2$ at each step. The complete Virasoro minimal modular invariants share the remarkable ADE structure as the simply-laced Lie algebras \cite{Cappelli:1986hf, Cappelli:1987xt}. We consider only diagonal Virasoro minimal models.

The Virasoro screening operators are 
\begin{equation}
    s^{\pm}(z) = \: : e^{ - i \alpha_{\pm} \alpha_{1} \cdot \phi } (z):\:  = \: : e^{ -i \sqrt{2} \alpha_{\pm} \phi  } (z):\:  .
\end{equation}
The free-field vertex operators are 
\begin{equation}
    V_{ ( r,s ) }(z) = \: : e^{i \sqrt{2} \alpha_{(r,s)} \phi }  : (z) , \quad \alpha_{(r,s)}= \frac{(r-1) \alpha_{+} + (s-1)\alpha_{-} }{2} ,
\end{equation}
and the dual vertex operators are 
\begin{equation}
    V_{(r,s)}^{\dag} =\: : e^{ -i \sqrt{2} ( \alpha_{0}+\alpha_{(r,s)}) } : (z) \: . 
\end{equation}
They both have the expected Virasoro minimal primary conformal weights 
\begin{equation}
    h_{(r,s)} =  \frac{ (rp' - sp)^{2} -(p-p')^{2} }{4pp'}. 
\end{equation}

\ 

\paragraph{The $\mathfrak{Vir}(5,2)$ minimal model.}

The simplest non-trivial Virasoro minimal model is the non-unitary $\mathfrak{Vir}(5,2)$ model. The $\mathfrak{Vir}(5,2)$ Kac-table is
\begin{equation}
    \big\{ L_{(1,1)} \equiv L_{(4,1)}, \: L_{(2,1)} \equiv L_{(3,1)} \big\},
\end{equation}
whose primary conformal weights are $h_{(1,1)}=0$ and $h_{(2,1)}=-\frac{1}{5}$. The $\mathfrak{Vir}(5,2)$ modular $S$ matrix is 
\begin{equation}
    S=  \begin{pmatrix}
        - \sqrt{\frac{5+\sqrt{5}}{10}}  &  \sqrt{\frac{5-\sqrt{5}}{10}} \\ \sqrt{\frac{5-\sqrt{5}}{10}} & \sqrt{\frac{5+\sqrt{5}}{10}}
    \end{pmatrix} .
\end{equation}
It is easy to verify that $S^{2}=C=I$. The non-zero fusion rules are
\begin{equation}
    L_{(2,1)} \times L_{(2,1)} = L_{(1,1)} +  L_{(2,1)}.
\end{equation}

Consider primary disk two-point functions $ \langle b_{(r,s)} \vert \Phi_{(2,1)} (z_{1},\widebar{z}_{1}) \Phi_{(2,1)} (z_{2},\widebar{z}_{2}) \vert 0 \rangle$. It consists of two types contributions: $ I_{(2,1)(2,1)}^{(1,1)} = \langle \langle L_{(1,1)} \vert \Phi_{(2,1)} (z_{1},\widebar{z}_{1}) \Phi_{(2,1)} (z_{2},\widebar{z}_{2}) \vert 0 \rangle$ and $ I_{(2,1)(2,1)}^{(2,1)} =  \langle \langle L_{(2,1)} \vert \Phi_{(2,1)} (z_{1},\widebar{z}_{1}) \Phi_{(2,1)} (z_{2},\widebar{z}_{2}) \vert 0 \rangle $. A free-field vertex operator realization of the contribution $I_{(2,1)(2,1)}^{(1,1)}$ is 
\begin{equation}    
  \oint dz^{(1)} \widebar{V}_{(2,1)}(\widebar{z}_{2}) \widebar{V}_{(2,1)}^{\dag}(\widebar{z}_{1})    s^{+} (z^{(1)}) V_{(2,1)}(z_{1}) V_{(2,1)}(z_{2})  , 
\end{equation}
whose result is 
\begin{equation*}
    \vert z_{1} \vert^{ \frac{4}{5}} (1-  \vert z_{1}\vert^{2})^{ \frac{2}{5} } \frac{\Gamma^{2}(\frac{3}{5})}{ \Gamma(\frac{6}{5})} \: _{1}F_{2} \Big[ \frac{3}{5} , \frac{4}{5} ; \frac{6}{5} ; \vert z_{1} \vert^{2} \Big]. 
\end{equation*}
A free-field vertex operator realization of the contribution $I_{(2,1)(2,1)}^{(2,1)}$ is 
\begin{equation}
   \oint d\widebar{z}^{(1)} \widebar{V}_{(2,1)}(\widebar{z}_{2}) \widebar{V}_{(2,1)}^{\dag}(\widebar{z}_{1})  \widebar{s}^{+} (\widebar{z}^{(1)})  V_{(2,1)}(z_{1})   V_{(2,1)}(z_{2}) ,
\end{equation}
whose result is
\begin{equation}
   \vert z_{1}\vert^{\frac{2}{5} }   (1-\vert z_{1}\vert^{2 })^{\frac{2}{5}} \frac{  \Gamma(\frac{3}{5}) \Gamma(\frac{1}{5})}{ \Gamma (\frac{4}{5})} \: _{1}F_{2}\Big[ \frac{3}{5},  \frac{2}{5}; \frac{4}{5} ; \vert z_{1} \vert^{2}\Big] .
\end{equation}

\ 

\paragraph{The Ising model $\mathfrak{Vir}(4,3)$.}

The simplest unitary Virasoro minimal model is the Ising model $\mathfrak{Vir}(4,3)$, whose disk two-point functions are given in \cite{Kawai:2002vd, Kawai:2002pz}. We perform the calculations independently. The Ising Kac-table is 
\begin{equation}
   \big\{ L_{(1,1)} \cong L_{(3,2)} , \: L_{(1,2)} \cong L_{(3,1)}  , \: L_{(2,2)}  \cong L_{(2,1)} \big\},
\end{equation}
whose primary conformal weights are $h_{(1,1)}=0$, $h_{(1,2)}=\frac{1}{2}$, and $h_{(2,2)} = \frac{1}{16}$. The Ising modular $S$ matrix is 
\begin{equation}
    S= \frac{1}{2}  \begin{pmatrix}
        1 & 1 & \sqrt{2}  \\  1 & 1 & -\sqrt{2} \\ \sqrt{2} & -\sqrt{2} & 0
    \end{pmatrix} . 
\end{equation}
It can be easily verified that $S^{2}=C=I$. The non-zero fusion rules are 
\begin{equation}
    L_{(2,1)} \times L_{(2,1)} = L_{(1,1)} + L_{(3,1)}, \quad L_{(2,1)} \times L_{(3,1)}  = L_{(2,1)}, \quad  L_{(1,2)} \times L_{(1,2)} = L_{(1,1)}.
\end{equation}

We show the calculations of disk two-point functions $\langle b_{(r,s)} \vert  \Phi_{(2,1)} (z_{1}, \widebar{z}_{1}) \Phi_{(2,1)}(z_{2}, \widebar{z}_{2}) \vert 0 \rangle $, which consist of two contributions $I_{(2,1) (2,1)}^{(1,1)}$ and $I_{(2,1)(2,1)}^{(3,1)}$. A free-field vertex operator realization of $I_{(2,1) (2,1)}^{(1,1)}$ is
\begin{equation}
   \oint dz^{(1)} \widebar{V}_{(2,1)}(\widebar{z}_{2}) \widebar{V}_{(2,1)}^{\dag}(\widebar{z}_{1})  s^{+}(z^{(1)}) V_{(2,1)} (z_{1})   V_{(2,1)}(z_{2})  ,
\end{equation}
whose result is
\begin{equation}
    \vert z_{1} \vert^{-\frac{1}{4}} (1- \vert z_{1} \vert^{2} )^{-\frac{1}{8}} \frac{\Gamma^{2}(\frac{1}{4})}{ \Gamma(\frac{1}{2})} \: _{1}F_{2} \Big[\frac{1}{4},- \frac{1}{4};\frac{1}{2} ; \vert z_{1} \vert^{2}\Big] .
\end{equation}
A free-field vertex operator realization of the contribution $I_{(2,1)(2,1)}^{(3,1)}$ 
\begin{equation}
    \oint d\widebar{z}^{(1)}\:  \widebar{V}_{(2,1)}( \widebar{z}_{2})  \widebar{V}^{\dag}_{(2,1)} (\widebar{z}_{1})  \widebar{s}^{+}(\widebar{z}^{(1)}) V_{(2,1)}(z_{1}) V_{(2,1)}(z_{2}) ,
\end{equation}
whose result is
\begin{equation}
     \vert z_{1} \vert^{\frac{3}{4}}  (1- \vert z_{1} \vert^{2} )^{-\frac{1}{8}} \frac{ \Gamma(\frac{1}{4}) \Gamma(\frac{5}{4}) }{\Gamma(\frac{3}{2})} \: _{1}F_{2} \Big[ \frac{1}{4}, \frac{3}{4} ; \frac{3}{2} ; \vert z_{1} \vert^{2} \Big] . 
\end{equation}

\ 

\paragraph{The Virasoro minimal model $\mathfrak{Vir}(5,3)$.}

The $\mathfrak{Vir}(5,3)$ Kac-table is
\begin{equation}
    \big\{  L_{(1,1)}  \cong L_{(4,2)} , \: L_{(2,1)}  \cong  L_{(3,2)} , \: L_{(4,1)} \cong  L_{(1,2)} , \: L_{(2,2)} \cong  L_{(3,1)} \big\} ,
\end{equation}
whose primary conformal weights are $h_{(1,1)}=1$, $h_{(2,1)}=-1/20 $, $h_{(4,1)}=3/4$, and $h_{(2,2)}=1/5$. The $\mathfrak{Vir}(5,3)$ modular $S$ matrix is 
\begin{equation}
    S=  \frac{1}{2} \begin{pmatrix}
        \sqrt{  (1+\frac{1}{\sqrt{5}})} &  \sqrt{  (1-\frac{1}{\sqrt{5}})}  & - \sqrt{  (1+\frac{1}{\sqrt{5}})} & -\sqrt{  (1-\frac{1}{\sqrt{5}})} \\  \sqrt{  (1-\frac{1}{\sqrt{5}})} &  \sqrt{  (1+\frac{1}{\sqrt{5}})} & \sqrt{  (1-\frac{1}{\sqrt{5}})} & \sqrt{  (1+\frac{1}{\sqrt{5}})} \\- \sqrt{  (1+\frac{1}{\sqrt{5}})}  & \sqrt{  (1-\frac{1}{\sqrt{5}})} & -\sqrt{  (1+\frac{1}{\sqrt{5}})} & \sqrt{  (1-\frac{1}{\sqrt{5}})} \\ -\sqrt{  (1-\frac{1}{\sqrt{5}})} & \sqrt{  (1+\frac{1}{\sqrt{5}})} &  \sqrt{  (1-\frac{1}{\sqrt{5}})} & - \sqrt{  (1+\frac{1}{\sqrt{5}})}
    \end{pmatrix}.
\end{equation}
It is easy to verify that $S^{2}=C=I$. The $\mathfrak{Vir}(5,3)$ fusion rules are 
\begin{equation*}
    L_{(2,1)} \times L_{(2,1)} = L_{(1,1)} + L_{(3,1)} , \quad  L_{(2,1)} \times L_{(2,2)} = L_{(1,2)} + L_{(2,1)},
\end{equation*}
\begin{equation}
    L_{(1,2)} \times L_{(1,2)} = L_{(1,1)} , \quad   L_{(2,2)} \times L_{(2,2)} = L_{(1,1)}+L_{(3,1)} .
\end{equation}

Consider disk correlation functions with Cardy boundary conditions $\langle b_{(r,s)} \vert  \Phi_{(2,1)} (z_{1} , \widebar{z}_{1}) \Phi_{(2,1)} (z_{2},\widebar{z}_{2}) \vert 0 \rangle $, which consists of two contributions $I_{(2,1)(2,1)}^{(1,1)}$ and $I_{(2,1)(2,1)}^{(3,1)}$. A free-field vertex operator realization of contribution $I_{(2,1)(2,1)}^{(1,1)}$ is 
\begin{equation}
    \oint dz^{(1)} \widebar{V}_{(2,1)}(\widebar{z}_{2}) \widebar{V}_{(2,1)}^{\dag}(\widebar{z}_{1}) s^{+}(z^{(1)}) V_{(2,1)}(z_{1})  V_{(2,1)}(z_{2})  ,
\end{equation}
whose result is 
\begin{equation}
     \vert z_{1} \vert^{ \frac{1}{5} }   (1-\vert z_{1} \vert^{2})^{ \frac{1}{10}} \frac{\Gamma^{2}(\frac{2}{5})}{\Gamma(\frac{4}{5})} \: _{1}F_{2}\Big[\frac{2}{5},\frac{1}{5}; \frac{4}{5}; \vert z_{1} \vert^{2}\Big] . 
\end{equation}
A free-field vertex operator realization of contribution $I_{(2,1)(2,1)}^{(3,1)}$ is 
\begin{equation}
    \oint d\widebar{z}^{(1)} \widebar{V}_{(2,1)}(\widebar{z}_{2}) \widebar{V}_{(2,1)}^{\dag}(\widebar{z}_{1})  \widebar{s}^{+}(\widebar{z}^{(1)}) V_{(2,1)}(z_{1}) V_{(2,1)}(z_{2})  ,
\end{equation}
whose result is 
\begin{equation}
     \vert z_{1}\vert^{\frac{3}{5}} (1- \vert z_{1}\vert^{2})^{\frac{1}{10}}  \frac{ \Gamma(\frac{2}{5}) \Gamma(\frac{4}{5})}{ \Gamma(\frac{6}{5}) } \: _{1}F_{2} \Big[ \frac{2}{5}, \frac{3}{5}; \frac{6}{5}; \vert z_{1} \vert^{2}\Big] . 
\end{equation}

\ 

\paragraph{Virasoro minimal model $\mathfrak{Vir}(5,4)$.}

The diagonal Virasoro minimal model $\mathfrak{Vir}(5,4)$ is a sub-theory of the diagonal $\mathcal{N}=1$ super-Virasoro minimal model $\mathfrak{SVir}^{1}(5,3)$. The Virasoro central charge of the $\mathfrak{Vir}(5,4)$ minimal algebra is $c_{\text{Vir}}=7/10$. The Kac-table of $\mathfrak{Vir}(5,4)$ is 
\begin{equation*}
    \big\{ L_{(1,1)} \equiv L_{(4,3)} , \quad L_{(2,1)} \equiv L_{(3,3)} , \quad L_{(3,1)} \equiv L_{(2,3)} ,  
\end{equation*}
\begin{equation}
    L_{(4,1)} \equiv L_{(1,3)} , \quad L_{(1,2)} \equiv L_{(4,2)} , \quad L_{(2,2)} \equiv L_{(3,2)} \big\} .
\end{equation}
The primary conformal weights of them are $h_{(1,1)}=0$, $h_{(2,1)}= 1/10$, $h_{(3,1)}=3/5$, $h_{(4,1)}=3/2$, $h_{(1,2)}= 7/16$, and $h_{(2,2)}=3/80$. The fact that $h_{(4,1)}=3/2$ indicates that the Virasoro chiral algebra in this theory can be extended to a $\mathcal{N}=1$ super-Virasoro algebra with the same central charge. The modular $S$ matrix is 
\begin{equation}
    S = \frac{1}{2\sqrt{2}} \begin{pmatrix}
      a &  b  &  b &  a & \sqrt{2}a &     \sqrt{ 2 } b \\  b  &  - a  & -a  &  b & -\sqrt{ 2 } b  & \sqrt{ 2 }a \\ b &  - a & - a &   b &  \sqrt{2} b& -\sqrt{2} a \\  a  & b & b   & a & - \sqrt{2 } a & -\sqrt{ 2 }b \\ \sqrt{2 } a & -\sqrt{ 2} b & \sqrt{ 2 }b & -\sqrt{ 2 }a & 0 &  0  \\   \sqrt{ 2 } b &  \sqrt{2} a& - \sqrt{ 2 }a & - \sqrt{2}b & 0 &  0 
    \end{pmatrix} ,
\end{equation}
where $a=  \sqrt{ \frac{5-\sqrt{5}}{5} } $, $b=  \sqrt{ \frac{5+\sqrt{5}}{5} } $, with $ S^{2}=C=I$. The detailed $\mathfrak{Vir}(5,4)$ fusion rules list is omitted, only the one related to the example considered is shown
\begin{equation}
    L_{(3,1)} \times L_{(3,1)} = L_{(1,1)} + L_{(3,1)} .
\end{equation}

Consider the example $\langle b_{(r,s)} \vert \Phi_{(3,1)} (z_{1}, \widebar{z}_{1}) \Phi_{(3,1)} (z_{2}, \widebar{z}_{2}) \vert 0 \rangle$, they consist of two components $I_{(3,1)(3,1)}^{(1,1)}$ and $I_{(3,1)(3,1)}^{(3,1)}$. The reason for considering this specific example is that, multiple contour integrals are necessary. A free-field realization of the contribution $I_{(3,1)(3,1)}^{(1,1)}$ is
\begin{equation}
    \oint dz^{(1)} \oint dz^{(2)} \widebar{V}_{(3,1)}(\widebar{z}_{2}) \widebar{V}^{\dag}_{(3,1)} (\widebar{z}_{1})  V_{(3,1)}(z_{1}) s^{+}(z^{(2)}) s^{+}(z^{(1)}) V_{(3,1)}(z_{2}) .
\end{equation}
The contribution  $I_{(3,1)(3,1)}^{(1,1)}$ is
\begin{equation*}
 \sum_{k_{i} \in \mathbb{N} } \vert z_{1} \vert^{ -\frac{12}{5}+ 2k_{2}}(1- \vert z_{1} \vert^{2} )^{-\frac{6}{5}} \frac{(-\frac{3}{5})_{\sum k_{i}} (\frac{8}{5})_{k_{1}} (-\frac{6}{5})_{k_{2}}  }{(-\frac{6}{5})_{\sum k_{i}} (k_{1})! (k_{2})!}  
\end{equation*}
\begin{equation}
 \frac{\Gamma(-\frac{3}{5})\Gamma(\frac{13}{5}) }{\Gamma(2)}  \frac{\Gamma(\frac{2}{5}+k_{1}) \Gamma(-\frac{3}{5}) }{\Gamma(-\frac{1}{5}+k_{1})} \: _{1}F_{2} \Big[ \frac{2}{5}+k_{1}, -\frac{6}{5}; -\frac{1}{5}+k_{1} ; \vert z_{1} \vert^{2} \Big]  \label{eq: V54 dis I1} 
\end{equation}
A free-field realization of the contribution $I_{(3,1)(3,1)}^{(3,1)}$ is 
\begin{equation}
    \oint d\widebar{z} \oint dz \widebar{V}_{(3,1)}(\widebar{z}_{2}) \widebar{V}_{(3,1)}^{\dag} (\widebar{z}_{1}) \widebar{s}^{+}(\widebar{z}) s^{+}(z) V_{(3,1)}(z_{1}) V_{(3,1)}(z_{2}) . 
\end{equation}
The contribution $I_{(3,1)(3,1)}^{(3,1)}$ is
\begin{equation*}
    \sum_{k \in \mathbb{N} } \vert z_{1} \vert^{-\frac{6}{5}+2k_{1}+2k_{2}} (1- \vert z_{1} \vert^{2} )^{-\frac{6}{5}}\frac{ (-\frac{3}{5})_{\sum k_{i}} (-\frac{3}{5})_{k_{1}}(-\frac{3}{5})_{k_{2}} }{(-\frac{6}{5})_{\sum k_{i}} (k_{1})! (k_{2})! } 
\end{equation*}
\begin{equation}
   \frac{\Gamma^{2}(-\frac{3}{5})}{\Gamma(-\frac{6}{5})} \frac{\Gamma(-\frac{3}{5}+k_{1}) \Gamma(\frac{11}{5}) }{\Gamma(\frac{8}{5}+k_{1})}   \: _{1}F_{2} \Big[ -\frac{3}{5}+k_{1} , \frac{8}{5}  ; \frac{8}{5} ; \vert z_{1}  \vert^{2} \Big] . \label{eq: V54 dis I2}
\end{equation}

\ 

\subsection{$\mathcal{W}_{3}(p,p')$ minimal models}

The $\mathcal{W}_{3} (p,p') $ algebra is the principal $\mathcal{W} \big[ \widehat{A}_{2}(k) \big] $ algebra, with $k = p/p' -3$. The Cartan matrix of $A_{2}$ is
\begin{equation}
    A= \begin{pmatrix}
        2 & -1 \\ -1 & 2
    \end{pmatrix} .
\end{equation}
The $A_{2}$ simple roots are $\alpha_{1}=[2,-1]$ and $\alpha_{2}=[-1,2]$. The Weyl group $W$ has $6$ elements, generated by the simple Weyl reflections $w_{1} = \begin{pmatrix}
    -1 & 0 \\ 1 & 1
\end{pmatrix} $ and $w_{2} = \begin{pmatrix}
    1 & 1 \\ 0 & -1
\end{pmatrix}$. The longest element has length $l(w_{l})=3$. The $A_{2}$ highest root is $\theta= \alpha_{1} + \alpha_{2} = \alpha_{1}^{\vee} +\alpha_{2}^{\vee} $. The $\widehat{A}_{2}$ Coxeter and the dual Coxeter numbers are $h= h^{\vee}=3$. The $\mathcal{W}_{3}(p,p')$ minimal Kac-table is
\begin{equation}
    \widehat{\Lambda}^{(+)} \in \widehat{P}_{+}^{p-3} , \quad   \widehat{\Lambda}^{(-)} \in \widehat{P}_{+}^{p'-3}   .
\end{equation}
The $\widehat{A}_{2}$ Weyl symmetries permute the Dynkin labels of $\widehat{\Lambda}^{(\pm)}$. 

\ 

\paragraph{The $\mathcal{W}_{3}(5,3)$ model.}

The simplest non-unitary $\mathcal{W}_{3}$ minimal model is the $\mathcal{W}_{3}(5,3)$ model. The only allowed $\Lambda^{(-)}$ is $[0,0]$, and there are a total of six choice of $\Lambda^{(+)}$. We label the $\mathcal{W}_{3}(5,3)$ irreducible modules by $\Lambda^{(+)}$ only, denoted by $L_{\Lambda^{(+)}}$. The $\mathcal{W}_{3}(5,3)$ minimal Kac-table is
\begin{equation}
    \Big\{  L_{[0,0]} \equiv L_{[2,0]} \equiv L_{[0,2]} ,\:  L_{[1,1]}\equiv L_{[1,0]} \equiv L_{[0,1]} \Big\}.
\end{equation}
Both the irreducible modules are self charge-conjugated. The primary conformal weights of the two modules are
\begin{equation}
    h_{[0,0]}=0 , \quad h_{[1,1]}=-\frac{1}{5} .
\end{equation}
The $\mathcal{W}_{3}(5,3)$ modular $S$ matrix is  
\begin{equation}
    S=  \begin{pmatrix}
       - \sqrt{\frac{1}{2} (1+\frac{1}{\sqrt{5}}) }  & \sqrt{\frac{1}{2} (1-\frac{1}{\sqrt{5}}) } \\  \sqrt{\frac{1}{2} (1-\frac{1}{\sqrt{5}}) } & \sqrt{\frac{1}{2} (1+\frac{1}{\sqrt{5}}) }
    \end{pmatrix} .
\end{equation}
It is easy to verify that $S^{2} =C=I$. The non-obvious non-zero fusion rule is
\begin{equation}
    L_{[1,1]} \times L_{[1,1]} = L_{[0,0]} + L_{[1,1]}  .
\end{equation}

Consider the disk two-point functions with Cardy boundary conditions $\langle b_{\Lambda^{(+)}} \vert \Phi_{[1,1]}(z_{1},\widebar{z}_{1}) \Phi_{[1,1]}(z_{2},\widebar{z}_{2}) \vert 0 \rangle $ in the diagonal $\mathcal{W}_{3}(5,3)$ model, which consists of two contributions $ I_{[1,1] [1,1]}^{[0,0]}$ and $I_{[1,1][1,1]}^{[1,1]}$. A free-field vertex operator realization of $I_{[1,1][1,1]}^{[0,0]}$ is 
\begin{equation}
   \oint dz^{(1)} \oint dz^{(2)} \widebar{V}_{[1,1]}(\widebar{z}_{2})  \widebar{V}^{\dag}_{[1,1]}(\widebar{z}_{1})  s_{1}^{+} (z^{(1)})   s_{2}^{+} (z^{(2)})  V_{[1,0]}(z_{1})   V_{[0,1]}(z_{2})  ,
\end{equation}
whose result is 
\begin{equation}
    \vert z_{1}\vert^{\frac{4}{5}}  (1- \vert z_{1}\vert^{2})^{\frac{1}{5}} \frac{\Gamma^2 (\frac{2}{5})}{\Gamma (\frac{4}{5})} \sum_{k\in \mathbb{N}}   \frac{(\frac{1}{5})_{k}(\frac{2}{5})_{k}  }{ k! (\frac{4}{5})_{k}} \frac{\Gamma(\frac{2}{5})\Gamma(\frac{4}{5}+k)  }{\Gamma(\frac{6}{5}+k) }  \: _{1}F_{2} \Big[ k+\frac{4}{5},\frac{1}{5}; k+\frac{6}{5};   \vert z_{1}\vert^{2}\Big] .
\end{equation}
A free-field vertex operator realization of the contribution is $I_{[1,1][1,1]}^{[1,1]}$
\begin{equation}
     \oint d\widebar{z}^{(1)}  \oint d\widebar{z}^{(2)}   \widebar{V}_{[1,1]} (\widebar{z}_{2})     \widebar{V}_{[1,1]}^{\dag} (\widebar{z}_{1})\widebar{s}^{+}_{1} (\widebar{z}^{(1)})  \widebar{s}^{+}_{2}(\widebar{z}^{(2)}) V_{[1,0]}(z_{1})  V_{[0,1]}(z_{2}) , 
\end{equation}
whose results is 
\begin{equation}
    \vert z_{1}\vert^{\frac{2}{5}} (1-\vert z_{1}\vert^{2})^{\frac{1}{5}}    \frac{\Gamma^{2}(\frac{2}{5})}{\Gamma(\frac{4}{5})} \sum_{k\in \mathbb{N}}\frac{ (\frac{1}{5})_{k}(\frac{2}{5})_{k} }{k! (\frac{4}{5})_{k}} \frac{ \Gamma(\frac{1}{5}+k)\Gamma(\frac{4}{5}) }{\Gamma(1+k)} \: _{1}F_{2} \Big[ \frac{1}{5}+k , \frac{1}{5}; 1+k ;\vert z_{1}\vert^{2} \Big]. 
\end{equation}

\ 

\paragraph{The $\mathcal{W}_{3}(5,4)$ minimal models.}

The simplest unitary $\mathcal{W}_{3}$ minimal model is the $\mathcal{W}_{3}(5,4)$ minimal model. The diagonal $\mathcal{W}_{3}(5,4)$ minimal model contains a non-diagonal Virasoro sub-theory, which is the three-state Potts model. Studying the boundary three-state Potts model using free-field approaches to $\mathcal{W}_{3}$ minimal models has been given in \cite{Caldeira:2003zz}. We consider the free-field approach to disk two-point functions of the diagonal and the charge-conjugated $\mathcal{W}_{3}(5,4)$ minimal models. The $\mathcal{W}_{3}(5,4)$ Kac-table is 
\begin{equation}
    \Lambda^{(+)} \in \big\{ [0,0] ,  [2,0] ,[0,2],[1,0],[1,1],[0,1] \big\} , \quad  \Lambda^{(-)} \in \big\{ [0,0] ,[1,0],[0,1] \big\} .
\end{equation}
The $\mathbb{Z}_{3}$ Weyl symmetry allows us to fix the choice of $\Lambda^{(-)}=[0,0]$, and label all irreducible modules by the six $\Lambda^{(+)}$. The modules $L_{[0,0]}$ and $L_{[1,1]}$ are self-conjugated, while $UL_{[2,0]} = L_{[0,2]} $ and $UL_{[1,0]} = L_{[0,1]} $. The primary conformal weights of them are 
\begin{equation}
    h_{[0,0]}=0, \: h_{[1,1]}=\frac{2}{5},\: h_{[2,0]}=h_{[0,2]}=\frac{2}{3} , \: h_{[1,0]}=h_{[0,1]}= \frac{1}{15},
\end{equation}
The modular $S$ matrix of this model is given by 
\begin{equation}
    S=  \frac{1}{ 10 \sqrt{3}}  \begin{pmatrix}
        c  &   d  &  c  &  c &   d &   d\\   d & - c & d  &  d &  -c  & -c \\  c  & d  &  a & a^{\ast}  &  b &  b^{\ast}  \\  c & d & a^{\ast} & a & b^{\ast} &  b \\ d & -c & b & b^{\ast}  & -a &  -a^{\ast} \\  d & -c & b^{\ast} & b  & -a^{\ast} & -a
    \end{pmatrix} ,
\end{equation}
where $a=  2i ( e^{\frac{2\pi i}{15}}-2e^{-\frac{4\pi i}{15}} -  e^{\frac{8\pi i}{15}}+2e^{\frac{14\pi i}{15}}  ) $, $b=2i ( 2e^{-\frac{2\pi i}{15}} + e^{\frac{4\pi i}{15}} -2  e^{-\frac{8\pi i}{15}}-e^{-\frac{14\pi i}{15}}  )$, $c=\sqrt{50-10\sqrt{5}}$, and $d=\sqrt{50+10\sqrt{5}}$. It is easy to verify that $S^{2}=C$. From which we can obtain six Cardy boundary states for both diagonal and charge conjugate models. Applying the Verlinde formula, we obtain the non-obvious non-zero fusion rules
\begin{equation*}
    L_{[1,1]} \times L_{[1,1]} = L_{[0,0]} + L_{[1,1]}  , \quad  L_{[0,1]} \times L_{[0,1]} = L_{[1,0]} , \quad L_{[1,0]} \times L_{[1,0]} = L_{[0,1]} ,
\end{equation*}
\begin{equation}
    L_{[2,0]} \times L_{[0,2]} = L_{[0,0]} + L_{[1,1]}  , \quad  L_{[0,2]} \times L_{[0,2]}=L_{[2,0]} , \quad L_{[2,0]} \times L_{[2,0]}= L_{[0,2]}  .
\end{equation}

\ 

We consider only the charge conjugated $\mathcal{W}_{3} [5,4]$ model. Consider disk correlation functions the charge conjugated model first. A free-field realization of the contribution $^{U}I_{[1,0][1,0]}^{[0,1]}$ is 
\begin{equation}
      \oint d\widebar{z} \oint dz  \widebar{V}_{[0,1]} (\widebar{z}_{2}) \widebar{V}_{[1,0]}^{\dag}(\widebar{z}_{1}) \widebar{s}_{2}^{+} ( \widebar{z})   s_{1}^{+} (z)   V_{[1,0]} (z_{1}) V_{[1,0]} (z_{2}) .
\end{equation}
The contribution $^{U}I_{[1,0][1,0]}^{[0,1]}$ is  
\begin{equation*}
    \sum_{k_{i} \in \mathbb{N}}  \vert z_{1} \vert^{-\frac{2}{15} +2k_{1}+2k_{2} } (1- \vert z_{1} \vert^{2} )^{-\frac{2}{15}}\frac{ (\frac{1}{5})_{\sum k_{i}} (\frac{4}{5})_{k_{1}} (-\frac{2}{15} )_{k_{2}}  }{(\frac{2}{5})_{\sum k_{i}} (k_{1})! (k_{2})! } 
\end{equation*}
\begin{equation}
    \frac{\Gamma^{2}(\frac{1}{5})}{\Gamma(\frac{2}{5})}  \frac{\Gamma(k_{1}+\frac{1}{5}) \Gamma(\frac{3}{5}) }{\Gamma(k_{1}+\frac{4}{5})} B\Big(\frac{1}{5}+k_{1}, \frac{4}{5}+k_{1} \Big). 
\end{equation}

\ 

The correlation functions $\langle b_{\Lambda^{(+)}}  \vert \Phi_{[1,1]}(z_{1},\widebar{z}_{1}) \Phi_{[1,1]}(z_{2},\widebar{z}_{2}) \vert 0 \rangle $ and  $\langle c  \vert \Phi_{[1,1]}(z_{1},\widebar{z}_{1}) \Phi_{[1,1]}(z_{2},\widebar{z}_{2}) \vert 0 \rangle $ are also considered. Two non-zero contributions exist for the correlation functions $I_{[1,1][1,1]}^{[0,0]}$ and $I_{[1,1][1,1]}^{[1,1]}$. The contribution $I_{[1,1][1,1]}^{[0,0]}$ has a free-field vertex operator realization 
\begin{equation*}
     \oint  dz^{(1)} \oint  dz^{(2)}   \oint  dz^{(3)}\oint  dz^{(4)} \widebar{V}_{[1,1]}(\widebar{z}_{2}) \widebar{V}_{[1,1]}^{\dag} (\widebar{z}_{1}) 
\end{equation*}
\begin{equation}
     s^{+}_{2} (z^{(4)}) s^{+}_{2} (z^{(3)}) s^{+}_{1} (z^{(2)})  s^{+}_{1} (z^{(1)})  V_{[1,1]}(z_{1})   V_{[1,1]}(z_{2})  .
\end{equation}
The disk contribution $I_{[1,1][1,1]}^{[0,0]}$ is
\begin{equation*}
    \sum_{k,l,m\in \mathbb{N}} \vert z_{1} \vert^{-\frac{8}{5} +2k_{2}+2k_{3} +2 k_{4}+2l_{2}  +2l_{3}  + 2m_{2} } (1-  \vert z_{1} \vert^{2} )^{-\frac{4}{5}} \frac{\Gamma(\frac{13}{5}) \Gamma(\frac{1}{5}) }{\Gamma(\frac{14}{5})} \frac{\Gamma(2)\Gamma(\frac{1}{5}+\sum_{i} k_{i})}{\Gamma(\frac{11}{5}+\sum_{i} k_{i}) } 
\end{equation*}
\begin{equation*}
   \frac{(\frac{13}{5})_{\sum_{k_{i}}} (\frac{4}{5})_{k_{1}}(\frac{4}{5})_{k_{2}}(\frac{4}{5})_{k_{3}}(-\frac{2}{5})_{k_{4}} }{(\frac{14}{5})_{\sum_{k_{i}}} (k_{1})! (k_{2})! (k_{3})! (k_{4})!  }   \frac{ (\frac{1}{5}+\sum_{i} k_{i})_{ \sum_{i} l_{i}}  (\frac{4}{5})_{l_{1}} (\frac{4}{5})_{l_{2}} (-\frac{2}{5})_{l_{3}} }{(\frac{11}{5}+\sum_{i} k_{i})_{ \sum_{i} l_{i}} (l_{1})! (l_{2})! (l_{3})! } 
\end{equation*}
\begin{equation}
  \frac{\Gamma(a')\Gamma(c'-a')}{\Gamma(c')}  \frac{  (a)_{\sum_{i} m_{i} } (\frac{4}{5})_{m_{1}} (-\frac{2}{5})_{m_{2}} }{ (c)_{\sum_{i} m_{i}} (m_{1})!(m_{2})!  } \frac{\Gamma(a)\Gamma(c-a)}{\Gamma(c)} \: _{1}F_{2}\big[ a , -\frac{2}{5} ; c ; \vert z_{1} \vert^{2} \big], 
\end{equation}
where $a' = \frac{3}{5} + k_{2} + k_{3} + k_{4}+ \sum l_{i} $, $c' = \frac{16}{5} + k_{2} + k_{3} + k_{4}+ \sum l_{i} $, $a=\frac{4}{5}+ k_{3} +k_{4} + l_{2}+l_{3}+m_{2}$, and $c=1+ k_{3} +k_{4} + l_{2}+l_{3}+m_{2}$.

A free-field realization of the contribution $I_{[1,1][1,1]}^{[1,1]}$ is 
\begin{equation*}
     \oint  dz^{(1)}  \oint  dz^{(2)}   \oint  d\widebar{z}^{(1)}  \oint   d\widebar{z}^{(2)}\widebar{V}_{[1,1]} (\widebar{z}_{2})  \widebar{V}_{[1,1]}^{\dag} (\widebar{z}_{1})
\end{equation*}
\begin{equation}
      \widebar{s}^{+}_{1} (\widebar{z}^{(1)}) \widebar{s}^{+}_{2} (\widebar{z}^{(2)})    s^{+}_{1} (z^{(1)})  s^{+}_{2} (z^{(2)})  V_{[1,1]}(z_{1})   V_{[1,1]}(z_{2})  .
\end{equation}
The disk contribution $I_{[1,1][1,1]}^{[0,0]}$ is
\begin{equation*}
    \sum_{k,l,m \in \mathbb{N}}  \vert z_{1}\vert^{-\frac{4}{5} + 2k_{2}+2k_{3}+2k_{4} +2m_{2} +2m_{3}+2l_{2} } (1-  \vert z_{1}\vert^{2})^{-\frac{4}{5}}  \frac{\Gamma^{2}(\frac{1}{5})}{\Gamma(\frac{2}{5})}  \frac{\Gamma(\frac{1}{5}+l_{2})\Gamma(\frac{1}{5})}{\Gamma(\frac{2}{5}+l_{2})} 
\end{equation*}
\begin{equation*}
  \frac{(\frac{1}{5})_{\sum_{i} k_{i}} (\frac{4}{5})_{k_{1}} (\frac{4}{5})_{k_{2}} (-\frac{8}{5})_{k_{3}} (-\frac{2}{5})_{k_{4}} }{(\frac{2}{5})_{\sum_{i} k_{i}} (k_{1})! (k_{2})!(k_{3})!(k_{4})! }  \frac{(-\frac{2}{5}+\sum_{i}k_{i}) _{\sum_{i} l_{i} }  (-\frac{8}{5})_{l_{1}} (\frac{4}{5})_{l_{2}} (-\frac{2}{5})_{l_{3}}  }{(-\frac{1}{5}+\sum_{i}k_{i}) _{\sum_{i} l_{i} } (l_{1})!(l_{2})!(l_{3})! } 
\end{equation*}
\begin{equation}
   \frac{ (\frac{1}{5}+l_{2})_{\sum_{i} m_{i} } (-\frac{2}{5})_{m_{1}} (\frac{4}{5})_{m_{2}} }{(\frac{2}{5}+l_{2})_{\sum_{i} m_{i} } (m_{1})! (m_{2})!  } \frac{\Gamma(\frac{1}{5}+m_{1})\Gamma(\frac{7}{5})}{\Gamma(\frac{8}{5}+m_{1})} \: _{1}F_{2} \Big[ \frac{1}{5}+m_{1} ,  \frac{4}{5} ; \frac{8}{5}+m_{1}; \vert z_{1} \vert^{2} \Big]
\end{equation}

\

\paragraph{The $\mathcal{W}_{3}(7,3)$ model.}

The Virasoro central charge of the $\mathcal{W}_{3}(7,3)$ algebra is $c_{\text{Vir}}= -\frac{114}{7}$. The allowed $\Lambda^{(-)}$ for the irreducible representations is $[0,0]$. The irreducible representations are denoted by $L_{\Lambda^{(+)}}$, $\widehat{\Lambda}^{(+)} \in \widehat{P}^{4}_{+} $. The non-degenerate $\mathcal{W}_{3}(7,3)$ Kac-table is
\begin{equation*}
    \{  L_{[0,0]} \equiv L_{[4,0]}  \equiv L_{[0,4]} , \quad L_{[1,1]} \equiv L_{[2,1]} \equiv L_{[1,2]} , \quad L_{[2,2]}  \equiv L_{[0,2]} \equiv L_{[2,0]} , 
\end{equation*}
\begin{equation}
     L_{[1,0]} \equiv L_{[3,1]} \equiv L_{[0,3]}  ,\quad L_{[0,1]}\equiv L_{[3,0]} \equiv L_{[1,3]}  \}.
\end{equation}
The irreducible modules $L_{[0,0]}$, $L_{[1,1]}$, and $L_{[2,2]}$ are self charge-conjugated modules, while the rest satisfy $UL_{[3,0]} = L_{[1,0]}$. Their primary conformal weights are 
\begin{equation}
    h_{[0,0]}=0 , \quad  h_{[1,1]}= -\frac{5}{7} , \quad h_{[2,2]}=-\frac{4}{7} , \quad h_{[1,0]} = h_{[0,1]} = -\frac{3}{7} .
\end{equation}
The $\mathcal{W}_{3}(7,3)$ modular $S$ matrix are 
\begin{equation}
   S=  \frac{1}{7} \begin{pmatrix}
        -c & d &  -e & -b & -b  \\  d  &  e  & c & -b & -b \\ -e  &  c &  d  &  b  &  b \\  -b & -b  &  b  & a & a^{\ast}   \\ -b & -b  &  b  & a^{\ast} &  a
    \end{pmatrix} , \quad S^{2}=C , 
\end{equation}
where $a= i (3- e^{\frac{6\pi i}{7}} -e^{-\frac{2\pi i}{7}}- e^{-\frac{4\pi i}{7}} ) $ and $b=2 (\sin{\frac{6\pi}{7}} - \sin{\frac{4\pi}{7}} -\sin{\frac{2\pi}{7}} )$, $c= 2 (\sin{\frac{2\pi}{7}} +2 \sin{\frac{6\pi}{7}} ) $, $d=-2 ( \sin{\frac{4\pi}{7}} - 2\sin{ \frac{2\pi}{7} })$, and $e=2( 2\sin{\frac{4\pi}{7}} +\sin{\frac{6\pi}{7}} )$. The fusion rules are 
\begin{equation*}
    L_{[1,1]} \times L_{[1,1]}  = L_{[0,0]} + 2L_{[1,1]} +L_{[2,2]} + L_{[1,0]} + L_{[0,1]} , 
\end{equation*}
\begin{equation*}
     L_{[1,1]} \times L_{[2,2]} =L_{[1,1]} +  L_{[2,2]} + L_{[1,0]} + L_{[0,1]} , 
\end{equation*}
\begin{equation*}
    L_{[1,1]} \times L_{[1,0]} = L_{[1,1]} + L_{[2,2]} + L_{[1,0]} , \quad L_{[1,1]} \times L_{[0,1]} = L_{[1,1]} + L_{[2,2]} + L_{[0,1]}  , 
\end{equation*}
\begin{equation}
    L_{[2,2]} \times L_{[2,2]} = L_{[0,0]}  + L_{[1,1]}  +  L_{[2,2]} , \quad L_{[1,0]} \times L_{[0,1]} =L_{[0,0]} + L_{[1,1]} .
\end{equation}

\

Disk two-point correlation functions in the charge-conjugated model is considered $_{I}\langle b_{\Lambda^{(+)}} \vert \Phi_{[1,0]}(z_{1},\widebar{z}_{1})  \Phi_{[0,1]}(z_{2},\widebar{z}_{2}) \vert 0 \rangle$, which contains contributions $I_{[1,0][0,1]}^{[0,0]}$ and $I_{[1,0][0,1]}^{[1,1]}$. A free-field realization of the contribution $I_{[1,0][0,1]}^{[0,0]}$ is
\begin{equation}
    \oint dz^{(1)}  \oint dz^{(2)} \widebar{V}_{[1,0]}  (\widebar{z}_{2}) \widebar{V}_{[1,0]}^{\dag}(\widebar{z}_{1})  s_{1}^{+} (z^{(1)}) s_{2}^{+} (z^{(2)}) V_{[1,0]} (z_{1}) V_{[0,1]}(z_{2})  . 
\end{equation}
The disk contribution $I_{[1,0][0,1]}^{[0,0]}$ is
\begin{equation}
    \sum_{k\in \mathbb{N}} \vert z_{1} \vert^{\frac{12}{7}+2k} (1-\vert z_{1} \vert^{2}  )^{\frac{6}{7}}\frac{(\frac{4}{7})_{k}}{(k)!}  \frac{\Gamma^{3}(\frac{4}{7}) \Gamma(\frac{8}{7}+k) }{\Gamma(\frac{8}{7})\Gamma(\frac{12}{7}+k)} \:   _{1}F_{2} \Big[ \frac{4}{7}  , \frac{5}{7} ;  (\frac{12}{7}+k)  ; \vert z_{1} \vert^{2} \Big]. 
\end{equation}
A free-field realization of the contribution $I_{[1,0][0,1]}^{[1,1]}$ is
\begin{equation}
    \oint dz^{(1)}  \oint dz^{(2)} \widebar{V}_{[1,0]}  (\widebar{z}_{2}) \widebar{V}_{[1,0]}^{\dag}(\widebar{z}_{1})  \widebar{s}_{1}^{+}(\widebar{z}_{1}) \widebar{s}_{2}^{+}(\widebar{z}_{2})  V_{[1,0]} (z_{1}) V_{[0,1]}(z_{2})  . 
\end{equation}
The final result is 
\begin{equation}
    \sum_{k \in \mathbb{N}} \vert z_{1} \vert^{\frac{2}{7} + k }  \frac{ (1)_{k} (\frac{8}{7})_{k}}{k! (\frac{11}{7})_{k} } \frac{\Gamma(k+\frac{8}{7})\Gamma(\frac{2}{7}) }{ \Gamma(k+\frac{10}{7}) } \: _{1}F_{2} \big[ k+\frac{8}{7}  , \frac{3}{7}  ; k+\frac{10}{7}  ; \vert z_{1} \vert^{2}  \big] .
\end{equation}

\ 

\subsection{The $\mathcal{W}_{4}(7,4)$ minimal model}

The $\mathcal{W}_{4}(p,p')$ algebra is the principal $\mathcal{W} \big[ \widehat{A}_{3} (k )\big] $ algebra at level $k= \frac{p}{p'}-4$. The $A_{3}$ Cartan matrix is
\begin{equation}
    A = \begin{pmatrix}
        2 & -1 & 0 \\ -1 & 2 & -1 \\ 0  & -1 & 2
    \end{pmatrix},
\end{equation}
with the simple roots and coroots being $\alpha_{1}= \alpha_{1}^{\vee} = [2,-1,0] $, $\alpha_{2}= \alpha_{2}^{\vee} = [-1,2,-1] $, and $\alpha_{3}= \alpha_{3}^{\vee} = [0,-1,2] $. The $A_{3}$ Weyl group has a total of $24$ elements, generated by three simple Weyl reflections $w_{1}  = \begin{pmatrix}
    -1 & 0 & 0 \\ 1 & 1 & 0 \\ 0 & 0 & 1
\end{pmatrix} $, $w_{2}= \begin{pmatrix}
    1 & 1 & 0 \\ 0 & -1 & 0 \\ 0 & 1 & 1
\end{pmatrix} $, and $w_{3} = \begin{pmatrix}
    1 & 0 & 0 \\ 0 & 1 & 1\\ 0 & 0 & -1
\end{pmatrix} $. The longest element has length $l(w_{l})=6$. The highest root of $A_{3}$ is $\theta=\alpha_{1}+\alpha_{2}+\alpha_{3}= \alpha_{1}^{\vee} + \alpha_{2}^{\vee} +\alpha_{3}^{\vee}$, hence the Coxeter and the dual Coxeter numbers of $\widehat{A}_{3}$ are $h=h^{\vee}=4$. $ \vert P/ Q^{\vee}\vert = \det A_{ij} = 4$. The Weyl symmetry acting on the Kac-table as a $\mathbb{Z}_{4}$ permutation group on $\widehat{\Lambda}^{(\pm)}$. The $\mathcal{W}_{4}(7,4)$ algebra has Virasoro central charge $c= -\frac{114}{7}$. The non-degenerate $\mathcal{W}_{4}(7,4)$ minimal Kac-table $\{ L_{\widehat{\Lambda}^{(+)}}  \}$ is 
\begin{equation}
\Big\{  L_{ [0,0,0]  }   ,\quad L_{  [0,1,0]  } , \quad L_{ [1,1,1] }, \quad  L_{ [1,0,0] }, \quad L_{ [0,0,1] }  \Big\},
\end{equation}
with $L_{ [0,0,0]  }$, $L_{  [0,1,0]  }$, and $L_{ [1,1,1] }$ being self charge conjugated modules, and $UL_{[1,0,0]} = L_{[0,0,1]}$. The primary conformal weights of them are 
\begin{equation}
    h_{[0,0,0]}=0 , \quad h_{[0,1,0]}=-\frac{4}{7} , \quad h_{[1,1,1]}= - \frac{5}{7} , \quad h_{[1,0,0]}= h_{[0,0,1]} = -\frac{3}{7} . 
\end{equation}
The $\mathcal{W}_{4}(7,4)$ minimal modular $S$ matrix is 
\begin{equation}
    S=  \frac{1}{\sqrt{7}} \begin{pmatrix}
       -\frac{2}{7} a &  -\frac{2}{7} b &  -\frac{2}{7}c & 1 & 1 \\  -\frac{2}{7} b  &  -\frac{2}{7} c  & \frac{2}{7}a  & -1  & -1 \\ -\frac{2}{7}c  &  \frac{2}{7}a  & \frac{2}{7}b &  1 & 1 \\ 1  & -1 & 1 & d  & d^{\ast} \\  1 & -1 & 1 & d^{\ast}  &  d
    \end{pmatrix} ,
\end{equation}
where $a=(-1 + 5\cos{\frac{2\pi}{7} } - 2\cos{\frac{4\pi}{7} }- 2\cos{\frac{6\pi}{7} }) $, $b=  (   1+ 2\cos{\frac{2\pi}{7} } + 2\cos{\frac{4\pi}{7} }-5 \cos{\frac{6\pi}{7} } )$, , $c= (-1 -2 \cos{\frac{2\pi}{7} } + 5 \cos{\frac{4\pi}{7} } - 2 \cos{\frac{6\pi}{7} } )$, and $d= \frac{1}{2} i(i+ \sqrt{7} )$. It is easy to verify that $S^{2}=C$. The Verlinde formula provides the following non-obvious fusion rules  
\begin{equation*}
    L_{[0,1,0]} \times L_{[0,1,0]} = L_{[0,0,0]} +L_{[0,1,0]} + L_{[1,1,1]} , 
\end{equation*}
\begin{equation*}
    L_{[0,1,0]} \times L_{[1,1,1]} = L_{[0,1,0]} + L_{[1,1,1]} + L_{[1,0,0]} + L_{[0,0,1]} 
\end{equation*}
\begin{equation*}
    L_{[1,1,1]} \times L_{[1,1,1]} = L_{[0,0,0]} + L_{[0,1,0]} +2 L_{[1,1,1]} + L_{[1,0,0]} + L_{[0,0,1]} ,
\end{equation*}
\begin{equation*}
    L_{[1,0,0]} \times L_{[1,0,0]}=L_{[0,0,1]} , \quad L_{[0,0,1]} \times L_{[0,0,1]} =L_{[1,0,0]} ,
\end{equation*}
\begin{equation}
    L_{[1,0,0]} \times L_{[0,0,1]} = L_{[0,0,0]} + L_{[1,1,1]}. 
\end{equation}

\ 

Consider the disk two-point functions $\langle b_{\Lambda^{(+)}} \vert \Phi_{[0,1,0]} (z_{1}, \widebar{z}_{1}) \Phi_{[0,1,0]} (z_{2}, \widebar{z}_{2})\vert 0 \rangle $, with a  total of three contributions $I_{[0,1,0][0,1,0]}^{[0,0,0]}$, $I_{[0,1,0][0,1,0]}^{[0,1,0]}$, and $I_{[0,1,0][0,1,0]}^{[1,1,1]}$. 

A free-field vertex operator realization of the contribution $I_{[0,1,0][0,1,0]}^{[0,0,0]}$ is 
\begin{equation*}
    \oint dz^{(1)}  \oint dz^{(2)}   \oint dz^{(3)} \oint dz^{(4)}\widebar{V}_{[0,1,0]}(\widebar{z}_{2}) \widebar{V}^{\dag}_{[0,1,0]}(\widebar{z}_{1})
\end{equation*}
\begin{equation}
    s_{3}^{+} (z^{(4)}) s_{1}^{+} (z^{(3)}) s_{2}^{+} (z^{(2)}) s_{2}^{+} (z^{(1)})  V_{[0,1,0]}(z_{1})  V_{[0,1,0]}(z_{2}).
\end{equation}
The final result is 
\begin{equation*}
    \sum_{k_{i}, l_{i}, m_{i} \in \mathbb{N}} \vert z_{1} \vert^{\frac{16}{7} + k_{3}+l_{2} +m_{2} } (1- \vert z_{1} \vert^{2} )^{\frac{8}{7}}  \frac{\Gamma(2+ \sum k_{i})\Gamma(\frac{3}{7})}{\Gamma(\frac{17}{7}+ \sum k_{i}) }
\end{equation*}
\begin{equation*}
    \frac{(1)_{\sum k_{i} } (\frac{4}{7})_{k_{1}} (\frac{4}{7})_{k_{2}} (\frac{6}{7})_{k_{3}} }{(k_{1})!(k_{2})!(k_{3})! (2)_{\sum k_{i} }}\frac{( 2+\sum k_{i} )_{\sum l_{i}} (\frac{4}{7})_{l_{1}} (\frac{6}{7})_{l_{2}} }{ (l_{1})! (l_{2})!  (\frac{17}{7}+\sum k_{i} )_{\sum l_{i}} }
\end{equation*}
\begin{equation}
    \frac{(a)_{\sum m_{i}} (\frac{4}{7})_{m_{1}} (\frac{2}{7})_{m_{2}} }{ (m_{1})! (m_{2})! (c)_{ \sum m_{i} } }  \frac{\Gamma( a') \Gamma(\frac{15}{7}) }{ \Gamma(c' ) }\frac{\Gamma(a) \Gamma(\frac{3}{7}) }{\Gamma(c)}  \:  _{1}F_{2} \big[ a , \frac{2}{7}  ; c ; \vert z_{1} \vert^{2} \big],
\end{equation}
where $a'=\frac{9}{7}+ k_{2} +k_{3}+l_{2}+l_{3}$, $c'= \frac{24}{7} + k_{2} +k_{3}+l_{2}+l_{3}$, $a=\frac{12}{7}+k_{3}+l_{2} +m_{2}$, and $c=\frac{15}{7}+k_{3}+l_{2} +m_{2}$.

\

A free-field vertex operator realization of the contribution $I_{[0,1,0][0,1,0]}^{[0,1,0]}$ is 
\begin{equation*}
     \oint d\widebar{z}^{(1)}  \oint d\widebar{z}^{(2)}  \oint d\widebar{z}^{(3)}  \oint d\widebar{z}^{(4)}  \widebar{V}_{[0,1,0]}^{\dag} (\widebar{z}_{2})  \widebar{V}_{[0,1,0]} (\widebar{z}_{1}) 
\end{equation*}
\begin{equation}
    \widebar{s}_{3}^{+} (\widebar{z}^{(4)}) \widebar{s}_{1}^{+} (\widebar{z}^{(3)})\widebar{s}_{2}^{+} (\widebar{z}^{(2)}) \widebar{s}_{2}^{+} (\widebar{z}^{(1)})  V_{[0,1,0]} (z_{1}) V_{[0,1,0]} (z_{2}).
\end{equation}
The disk contribution $I_{[0,1,0][0,1,0]}^{[0,1,0]}$ is 
\begin{equation*}
    \sum_{k_{i},l_{i},m_{i}\in \mathbb{N}} \vert z_{1}\vert^{ \frac{8}{7}+ 2m_{2}} (1-\vert z_{1}\vert^{2})^{ \frac{4}{7}} \frac{\Gamma (\frac{1}{7}) }{\Gamma (\frac{8}{7})} \frac{ (\frac{1}{7})_{ \sum k_{i}} (\frac{4}{7})_{k_{1}} (\frac{4}{7})_{k_{2}} }{k_{1}! k_{2}! (\frac{8}{7})_{ \sum k_{i}} } \frac{\Gamma(a') \Gamma(\frac{3}{7})}{ \Gamma(c')} \frac{ (a')_{l} (\frac{4}{7})_{l} }{ l! (c')_{l} } \frac{ \Gamma( a ) \Gamma(\frac{15}{7}) }{\Gamma( c )}  
\end{equation*}
\begin{equation}
    \frac{ (a)_{\sum m_{i} }  (\frac{4}{7})_{ m_{1} }  (\frac{4}{7})_{ m_{2} }  }{ m_{1}! m_{2}! (c)_{\sum m_{i} }}   \frac{ \Gamma(\frac{4}{7} + \sum m_{i}) \Gamma(\frac{3}{7}) }{ \Gamma (1 + \sum m_{i}) }  \: _{1}F_{2}\Big[ \frac{4}{7} + \sum m_{i} , \frac{4}{7}  ; 1+ \sum m_{i} ; \vert z_{1}\vert^{2} \Big]  . 
\end{equation}
The parameters are $a'=\frac{2}{7} +  \sum k_{i} $ and $c' = \frac{5}{7} +\sum k_{i}$, $a=-\frac{1}{7} +k_{2} +l $, $c=2 +k_{2} +l$.

\

A free-field vertex operator realization of contribution $I_{[0,1,0][0,1,0]}^{[1,1,1]}$ is
\begin{equation*}
    \oint d\widebar{z}^{3} \oint d\widebar{z}^{2} \oint d\widebar{z}^{1} \oint dz^{1} \widebar{V}_{[0,1,0]} (\widebar{z}_{2}) \widebar{V}^{\dag}_{[0,1,0]} (\widebar{z}_{1}) 
\end{equation*}
\begin{equation}
    \widebar{s}_{1}^{+}(\widebar{z}^{1}) \widebar{s}_{2}^{+}(\widebar{z}^{2}) \widebar{s}_{3}^{+}(\widebar{z}^{3}) s_{2}^{+} (z^{1}) V_{[0,1,0]}(z_{1}) V_{[0,1,0]} (z_{2})
\end{equation}
The disk contribution $I_{[0,1,0][0,1,0]}^{[1,1,1]}$ is
\begin{equation*}
   \sum_{k,l,m\in \mathbb{N}} \vert z_{1} \vert^{\frac{6}{7} + 2\sum_{i} k_{i}+l  }  (1-\vert z_{1} \vert^{2})^{\frac{8}{7}}\frac{\Gamma^{2}(\frac{3}{7})}{\Gamma(\frac{6}{7})} \frac{ (\frac{3}{7})_{\sum_{i} k_{i} } (\frac{4}{7})_{k_{1}} (-\frac{8}{7})_{k_{2}}  (\frac{4}{7})_{k_{3}}(\frac{2}{7})_{k_{4}} }{ (\frac{6}{7})_{\sum_{i} k_{i} } (k_{1})!(k_{2})!(k_{3})!(k_{4})!}
\end{equation*}
\begin{equation*}
    \frac{\Gamma(k_{1}+1) \Gamma(\frac{3}{7}) }{\Gamma(k_{1}+\frac{10}{7})} \frac{(k+1)_{l} (\frac{6}{7})_{l} }{ (k+\frac{10}{7})_{l} l!} \frac{\Gamma(\frac{6}{7}+k_{2}+l) \Gamma(\frac{3}{7}) }{\Gamma(\frac{9}{7}+k_{2}+l)}\frac{(\frac{6}{7}+k_{2}+l)_{\sum_{i} m_{i} } (\frac{2}{7})_{m_{1}}  (\frac{4}{7})_{m_{1}}  }{  (\frac{9}{7}+k_{2}+l)_{\sum_{i} m_{i} }   (m_{1})!(m_{2})! } 
\end{equation*}
\begin{equation}
    \frac{\Gamma(\frac{1}{7}) \Gamma(\frac{9}{7}+k_{2}+l+m_{1}) }{\Gamma(\frac{10}{7}+k_{2}+l+m_{1})} \: B\Big[(\frac{9}{7}+k_{2}+l+m_{1}),(\frac{10}{7}+k_{2}+l+m_{1}) \Big].
\end{equation}

\

\subsection{The $\mathcal{W}\big[ \widehat{B}_{2} \big](4,5)$ minimal model}
\label{subsec: WB2}

From now, we consider principal QDS $\mathcal{W}\big[ \widehat{g} \big](p,p')$ minimal models, with  non-simply-laced simple Lie algebra $g$. The simplest non-simply-laced Lie algebra is the $g=B_{2}\cong C_{2}$ algebra. To obtain the information in the simplest $\mathcal{W}\big[ \widehat{B}_{2} \big](p,p')$ minimal model, information on the $B_{2}$ algebra is reviewed.

The $B_{2}$ Cartan matrix is 
\begin{equation}
    A= \begin{pmatrix}
        2 & -2 \\ -1 & 2
    \end{pmatrix},
\end{equation}
where the simple roots and simple coroots are given by $\alpha_{1} = \alpha_{1}^{\vee} =[2,-2]$ and $\alpha_{2}= \frac{1}{2} \alpha_{2}^{\vee} =[-1,2]$. From which, we conclude $r^{\vee} =2$. The Weyl vector and the dual Weyl vectors of the $B_{2}$ algebra are $\rho=[1,1]$ and $\rho^{\vee}= [1,2]$, respectively. The Weyl group of $B_{2}$ contains eight elements, generated by the two simple Weyl reflections $w_{1}=\begin{pmatrix}
        -1 & 0 \\ 2 & 1
    \end{pmatrix} $, and $w_{2}= \begin{pmatrix}
        1 & 1 \\ 0 & -1
\end{pmatrix}$. The longest Weyl element has length $l(w_{l})=4$ and the charge conjugation is trivial $U=-w_{l}=I$. The $B_{2}$ highest root $\theta$ decomposes as $\theta =\alpha_{1}+2\alpha_{2} =\alpha_{1}^{\vee} + \alpha_{2}^{\vee} $. The Coxeter and dual Coxeter numbers of $\widehat{B}_{2}$ are $h=4$ and $h^{\vee} =3$, respectively. The zeroth Dynkin label of the $\widehat{B}_{2}$ weight vectors are related to the level $k$ and the other Dynkin labels by
\begin{equation}
    \lambda_{0} = k - \lambda_{1} - \lambda_{2} .
\end{equation}
The $\mathcal{W}\big[\widehat{B}_{2} \big](p,p')$ Virasoro central charge is given by
\begin{equation}
    c= 86 - \frac{60p}{p'} - \frac{30p'}{p}, 
\end{equation}
and the $\mathcal{W}\big[\widehat{B}_{2} \big](p,p')$ Kac-table is 
\begin{equation}
    \widehat{\Lambda}^{(+)}\in\widehat{P}_{+}^{ p-3} , \quad \widehat{\Lambda}^{(-)}\in \widehat{P}_{+}^{\vee \: p'-4} . 
\end{equation}

\ 

The simplest non-trivial $\mathcal{W}\big[\widehat{B}_{2} \big](p,p')$ minimal model is the $\mathcal{W} \big[ \widehat{B}_{2} \big] (4,5)$ Virasoro minimal model, whose Virasoro central charge is $ c_{\text{Vir}}= \frac{1}{2}$. The choices of the $B_{2}$ weights are 
\begin{equation}
    \Lambda^{(+)} \in \big\{ [0,0] , [1,0] , [0,1] \big\} , \quad  \Lambda^{(-)} \in \big\{ [0,0] , [1,0]  \big\} .
\end{equation}
The Weyl symmetry identifies the following representations 
\begin{equation}
    L_{([0,0],[0,0])}  \equiv L_{([1,0],[1,0])} , \quad   L_{([1,0],[0,0])}  \equiv L_{([0,0],[1,0])} , \quad  L_{([0,1],[0,0])}  \equiv L_{([0,1],[1,0])}.
\end{equation}
In the squeals, the representations are label by $\Lambda^{(+)}$. The primary conformal weights of the modules are 
\begin{equation}
    h_{([0,0],[0,0])}  =0 , \quad h_{([1,0],[0,0])} = \frac{1}{2}   , \quad h_{([0,1],[0,0])}  = \frac{1}{16} .
\end{equation}
The $\mathcal{W} \big[ \widehat{B}_{2} \big] (4,5)$ modular $S$ matrix is given by
\begin{equation}
   S= \frac{1}{2} \begin{pmatrix}
        1  & 1  & -\sqrt{2} \\ 1  &  1  & \sqrt{2}  \\  -\sqrt{2}& \sqrt{2}  &  0 
    \end{pmatrix} , \quad S^{2}=C=I .
\end{equation}
The $\mathcal{W} \big[ \widehat{B}_{2} \big] (4,5)$ fusion rules are 
\begin{equation}
    L_{[1,0]} \times L_{[1,0]} = L_{[0,0]} , \quad L_{[0,1]} \times L_{[0,1]} = L_{[0,0]} + L_{[1,0]} , \quad L_{[0,1]} \times L_{[1,0]} = L_{[0,1]} .
\end{equation}

\

The disk correlation functions $\langle b \vert \Phi_{[1,0]} (z_{1},\widebar{z}_{1})  \Phi_{[1,0]} (z_{2},\widebar{z}_{2}) \vert 0 \rangle $. All correlation functions contain only one contribution $I_{[1,0][1,0]}^{[0,0]}$. A free-field realization of the contribution $I_{[1,0][1,0]}^{[0,0]}$ is 
\begin{equation*}
    \oint dz^{1} \oint dz^{2} \oint dz^{3} \oint dz^{4} \widebar{V}_{[1,0]} (\widebar{z}_{2}) \widebar{V}_{[1,0]}^{\dag}(\widebar{z}_{1})
\end{equation*}
\begin{equation}
      V_{[1,0]} (z_{1}) s_{1}^{+} (z^{1}) s_{1}^{+} (z^{2}) s_{2}^{+} (z^{3}) s_{2}^{+} (z^{4})  V_{[1,0]} (z_{2})  .
\end{equation}
The result $I_{[1,0][1,0]}^{[0,0]}$ is given by
\begin{equation*}
   \sum_{k,l,m\in \mathbb{N}} \vert z_{1}\vert^{-2+2 (k_{2}+k_{3}+l_{2}+m_{2}) }  (1- \vert z_{1} \vert^{2} )^{-1} \frac{\Gamma(\frac{9}{4})}{\Gamma(\frac{13}{4})}\frac{(\frac{9}{4})_{\sum_{i} k_{i} } (\frac{5}{4})_{k_{1}} (\frac{5}{4})_{k_{2}} (\frac{3}{4})_{k_{3}}  }{ (\frac{13}{4})_{\sum_{i} k_{i} } (k_{1})!(k_{2})!(k_{3})!}
\end{equation*}
\begin{equation*}
      \frac{ \Gamma(\frac{13}{4}+\sum_{i} k_{i} ) \Gamma(-\frac{1}{4})}{\Gamma (3+\sum_{i} k_{i} )  }\frac{\Gamma(a') \Gamma(\frac{7}{2})}{\Gamma(c')}\frac{(\frac{13}{4}+\sum_{i} k_{i} )_{\sum_{i} l_{i}}  (\frac{5}{4})_{l_{1}} (\frac{3}{4})_{l_{2}} }{ (3+\sum_{i} k_{i} )_{\sum_{i}l_{i} } (l_{1})!(l_{2})! } 
\end{equation*}
\begin{equation}
     \frac{(a')_{\sum_{i} m_{i} }  (\frac{5}{4})_{m_{1}} (-\frac{7}{4})_{m_{2}}  }{(c')_{\sum_{i} m_{i}
     } (m_{1})!(m_{2})! } \frac{\Gamma(-\frac{1}{4}) \Gamma(a) }{\Gamma(c)}\: _{1}F_{2} \Big[ a , -\frac{7}{4}  ;  c ; \vert z_{1} \vert \Big] ,
\end{equation}
where $a'=\frac{7}{4}+k_{2}+k_{3}+l_{1}+l_{2}$, $c' = \frac{21}{4}+k_{2}+k_{3}+l_{1}+l_{2}$, $a=\frac{3}{2} +k_{3}+ l_{2}+m_{2} $, and $c=\frac{5}{4} +k_{3}+ l_{2}+m_{2}$.

\ 

\subsection{The $\mathcal{W}\big[ \widehat{G}_{2} \big](6,7)$ minimal model}

Finally, we consider $\mathcal{W}\big[ \widehat{G}_{2} \big](6,7)$ minimal model. The $G_{2}$ Cartan matrix is 
\begin{equation}
    A = \begin{pmatrix}
        2 & -3 \\ -1 & 2
    \end{pmatrix}.
\end{equation}
The simple roots and coroots are $\alpha_{1} = \alpha_{1}^{\vee}=[2,-3]$ and $\alpha_{2} = \frac{1}{3} \alpha_{2}^{\vee} =[-1,2]$. Hence, $r^{\vee}=3$, and the $G_{2}$ Weyl vector and dual Weyl vector are $\rho=[1,1]$ and $\rho^{\vee} = [1,3]$. The highest root $\theta = 2\alpha_{1} +3\alpha_{2} =2\alpha_{1}^{\vee}+\alpha_{2}^{\vee} $. The Coxeter and the dual of Coxeter number of $\widehat{G}_{2}$ are $h=6$ and $h^{\vee} = 4$, respectively. $\vert P / Q^{\vee} \vert = \det(\alpha_{i}^{\vee} , \alpha_{j}^{\vee}) = 3 $. The $G_{2}$ Weyl group has a total of 12 elements, generated by the two simple Weyl reflections $w_{1}=\begin{pmatrix}
        -1 & 0 \\ 3 & 1
    \end{pmatrix} $ and $w_{2} =\begin{pmatrix}
        1 & 1 \\ 0 & -1
\end{pmatrix}  $. The longest Weyl element has length $l(w_{l}) =6$. The zeroth Dynkin label of a $\widehat{G}_{2}$ weight vector $\widehat{\lambda}$ is given by
\begin{equation}
    \lambda_{0} =k - 2\lambda_{1} -\lambda_{2}. 
\end{equation}
Hence, the Kac table of the $\mathcal{W}\big[ \widehat{G}_{2} \big](6,7)$ minimal model is
\begin{equation}
    \Lambda^{(+)} \in  \big\{ [0,0] , [1,0] , [0,1] , [0,2] \big\} , \quad \Lambda^{(-)}= [0,0].
\end{equation}
We label them using $\Lambda^{(+)}$, and their primary conformal weights are given by
\begin{equation}
    h_{[0,0]}=0 ,  \quad h_{[1,0]} = -\frac{1}{3}, \quad h_{[0,1]}= -\frac{2}{3} ,\quad h_{[0,2]} = -\frac{5}{9} .
\end{equation}
The $\mathcal{W}\big[ \widehat{G}_{2} \big](6,7)$ modular $S$ matrix is 
\begin{equation}
   S= \frac{1}{3\sqrt{3}} \begin{pmatrix}
         a &   b   &  c &  -3  \\  b  &  -c &  -a & -3 \\ c  & -a   &  b &  3 \\  -3 & -3  &  3 &  0
   \end{pmatrix} ,
\end{equation}
$a=(\cos{\frac{8\pi}{9}}+ \cos{\frac{5\pi}{9}} -\cos{\frac{4\pi }{9}} - \cos{\frac{\pi}{9}} )$, $b=(\cos{\frac{\pi }{9}} + \cos{\frac{2\pi}{9}} - \cos{\frac{7\pi }{9}}-\cos{\frac{8\pi }{9}}) $, $c=(\cos{\frac{2\pi}{9}}-\cos{\frac{4\pi}{9}} + \cos{\frac{5\pi}{9}}-\cos{\frac{7\pi}{9}}) $. It is easy to verify that $S^{2} =C =I $. Applying the Verlinde formula to this model, we obtain the non-obvious non-zero fusion rules 
\begin{equation*}
    L_{[1,0]} \times L_{[1,0]} = L_{[0,0]} +L_{[0,2]} , \quad L_{[1,0]} \times L_{[0,1]} = L_{[0,1]} + L_{[0,2]} ,
\end{equation*}
\begin{equation}
    L_{[0,1]} \times L_{[0,1]} = L_{[0,0]}+L_{[1,0]} + L_{[0,1]} +L_{[0,2]} , \quad  L_{[0,2]} \times L_{[0,2]} = L_{[0,0]} + L_{[0,1]} .
\end{equation}

\

Consider the disk two-point functions $\langle b_{\Lambda^{(+)}} \vert  \Phi_{[1,0]}(z_{1}, \widebar{z}_{1}) \Phi_{[1,0]}(z_{2}, \widebar{z}_{2}) \vert 0 \rangle $, which consist of two contributions $I_{[1,0],[1,0]}^{[0,0]}$ and $I_{[1,0],[1,0]}^{0,2}$. The contribution $ I_{[1,0][0,1]}^{[0,1]} $ is realized by 
\begin{equation*}
     \oint d\widebar{z}^{(1)} \oint dz^{(5)} \oint dz^{(4)} \oint dz^{(3)} \oint dz^{(2)} \oint dz^{(1)}  \widebar{V}_{[0,1]} (\widebar{z}_{2}) \widebar{V}_{[1,0]}^{\dag} (\widebar{z}_{1}) \widebar{s}_{2}^{+} (\widebar{z}^{(1)})
\end{equation*}
\begin{equation*}
     s^{+}_{1} (z^{(5)})  s^{+}_{1} (z^{(2)})   s^{+}_{2} (z^{(3)})   s^{+}_{2} (z^{(4)}) s^{+}_{2} (z^{(5)})   V_{[1,0]} (z_{1}) V_{[0,1]} (z_{2})  .
\end{equation*}
The integral result is 
\begin{equation*}
   \sum_{k,l,m,n,o} \vert z_{1} \vert^{ \frac{2}{3} + 2 (\sum_{i} k_{i}) + 2l_{5} + 2m_{4} +2 n_{2} +2o } (1- \vert z_{1} \vert^{2} )^{\frac{2}{3}}  \frac{\Gamma(\frac{11}{18})\Gamma(-\frac{2}{9})}{\Gamma(\frac{7}{18})}\frac{\Gamma(k_{1}+1)\Gamma(\frac{10}{3})}{\Gamma(k_{1}+\frac{13}{3})}  
\end{equation*}
\begin{equation*}
    \frac{ (\frac{11}{18})_{\sum k_{i}} ( \frac{7}{6})_{k_{1}} ( \frac{7}{6})_{k_{2}} ( -\frac{7}{9})_{k_{3}}( -\frac{7}{9})_{k_{4}} ( -\frac{7}{9})_{k_{5}} }{k_{1}!k_{2}!k_{3}! k_{4}! k_{5}! (\frac{7}{18})_{\sum k_{i}} } \frac{ (k_{1}+1)_{\sum l_{i}} (\frac{7}{6})_{l_{1}}(\frac{7}{6})_{l_{2}} (\frac{7}{6})_{l_{3}}(\frac{7}{6})_{l_{4}} (-\frac{3}{2})_{l_{5}} }{ l_{1}! l_{2}! l_{3} ! l_{4}! l_{5}! (k_{1}+\frac{13}{3})_{\sum l_{i}} } 
\end{equation*}
\begin{equation*}
    \frac{\Gamma(k_{1}+k_{2}+\sum l_{i} + \frac{13}{3} ) \Gamma(-\frac{1}{6}) }{\Gamma(k_{1}+k_{2}+\sum l_{i} + \frac{25}{6})}  \frac{ (k_{1}+k_{2}+\sum l_{i} + \frac{13}{3} )_{\sum m_{i} }  (\frac{7}{6})_{m_{1}} (\frac{7}{6})_{m_{2}} (\frac{7}{6})_{m_{3}} (-\frac{3}{2})_{m_{4}} }{(m_{1})!(m_{2})!(m_{3})!(m_{4})! (k_{1}+k_{2}+\sum l_{i} + \frac{25}{6} )_{\sum m_{i} }  } 
\end{equation*}
\begin{equation}
  \frac{\Gamma(a'')\Gamma(\frac{16}{9})}{\Gamma(c'')}  \frac{(a'')_{\sum n_{i} } ( -\frac{7}{9} )_{n_{1}} (\frac{11}{9})_{n_{2}}  }{(n_{1})!(n_{2})! (c'')_{\sum n_{i} } } \frac{\Gamma(a')\Gamma(\frac{10}{3})}{\Gamma(c')}  \frac{a_{o} (\frac{11}{9})_{o} }{(o)! c_{o} } \frac{\Gamma(a)}{\Gamma(c)} \: _{1}F_{2} \big( a  , \frac{11}{9}  ; c ; \vert z_{1} \vert^{2}  \big)
\end{equation}
where $a''= k_{1}+k_{2}+k_{3} +l_{2}+l_{3}+l_{4}+l_{5} +\sum m_{i}  $, $c''= a'' + \frac{16}{9}$, $a'= k_{1}+k_{2}+k_{3}+k_{4}+ m_{2}+m_{3}+m_{4}+l_{2}+l_{3}+l_{4}+l_{5}+  \sum n_{i} +\frac{5}{3} $, $c'=a'+ \frac{10}{3} $, $a= \sum_{i} k_{i} +l_{2}+l_{3}+l_{4}+l_{5} + m_{2}+m_{3}+m_{4} +n_{2} +o +\frac{83}{18} $, and $c=a+1$.

\

Consider the contribution $I_{[1,0][0,1]}^{[0,2]}$. A free-field realization of the contribution is 
\begin{equation*}
   \oint dz^{1} \oint dz^{2} s_{2}^{+}(z^{2}) s_{1}^{+}(z^{1}) V_{[1,0]}(z_{1}) V_{[0,1]}(z_{2})\oint d\widebar{z}^{1}
\end{equation*}
\begin{equation}
   \oint d\widebar{z}^{2} \oint d\widebar{z}^{3} \oint d\widebar{z}^{4}  \widebar{s}_{2}(\widebar{z}^{4})\widebar{s}_{2}(\widebar{z}^{3}) \widebar{s}_{2}(\widebar{z}^{2}) \widebar{s}_{1}(\widebar{z}^{1}) \widebar{V}_{[1,0]}^{\dag} (\widebar{z}_{1}) \widebar{V}_{[0,1]}(\widebar{z}_{2}) .
\end{equation}
The integral result is 
\begin{equation*}
    \sum_{k,l,m,n,o}  \vert z_{1} \vert^{\frac{8}{9} +2\sum k_{i} + 2\sum l_{i} + 2m_{3} +2n_{2}+2o }  (1-\vert z_{1} \vert^{2} )^{\frac{2}{3}} \frac{\Gamma(\frac{11}{18}) \Gamma(-\frac{1}{6}) }{\Gamma(\frac{4}{9})} \frac{\Gamma(\frac{4}{9}+ \sum k_{i} ) \Gamma(-\frac{1}{6})}{\Gamma(\frac{5}{18}+ \sum k_{i} )}
\end{equation*}
\begin{equation*}
    \frac{(\frac{11}{18})_{\sum k_{i}} (-\frac{7}{9})_{k_{1}} (-\frac{7}{9})_{k_{2}} (-\frac{7}{9})_{k_{3}} (\frac{7}{6})_{k_{4}} ( \frac{11}{9})_{k_{5}} }{  (k_{1})! (k_{2})! (k_{3})! (k_{4})! (k_{5})! (\frac{4}{9})_{\sum k_{i}}  }   \frac{ (\frac{4}{9})_{\sum l_{i}} (\frac{7}{6})_{l_{1}} (\frac{7}{6})_{l_{2}} (\frac{7}{6})_{l_{3}} (-\frac{7}{3})_{l_{4}} }{ (l_{1})!(l_{2})! (l_{3})! (l_{4})! (l_{5})! (\frac{5}{18})_{\sum l_{i}}  }  
\end{equation*}
\begin{equation*}
    \frac{\Gamma(k_{1}+l_{1}+\frac{11}{18})\Gamma(\frac{16}{9})}{\Gamma(k_{1}+l_{1}+\frac{43}{18})} \frac{(k_{1}+l_{1}+\frac{11}{18})_{ \sum m_{i}} (-\frac{7}{9})_{m_{1}} (\frac{7}{6})_{m_{2}} (\frac{11}{9})_{m_{3}} }{ (m_{1})!(m_{2})! (m_{3})! (k_{1}+l_{1}+\frac{43}{18})_{\sum m_{i} } }  \frac{\Gamma(a'' ) \Gamma(\frac{16}{9}) }{\Gamma(c'' ) } 
\end{equation*}
\begin{equation}
    \frac{(a'')_{\sum n_{i}}  (\frac{7}{6})_{n_{1}} (\frac{11}{9})_{n_{2}} }{(n_{1})! (n_{2})! (c'')_{\sum n_{i}} }  \frac{\Gamma(a')}{\Gamma(c')}\frac{(a')_{o} (\frac{11}{9})_{o} }{o! (c')_{o} } \frac{\Gamma(a) \Gamma (\frac{5}{2}) }{\Gamma(c)} \: _{1}F_{2} \big( a , \frac{7}{6}  ; c ; \vert z_{1} \vert^{2} \big) ,
\end{equation}
where $a''=k_{1}+k_{2}+k_{3}+l_{1}+l_{2}+l_{3}+m_{4}+\frac{25}{9}$, $c''=a''+\frac{16}{9}$, $a'=k_{1}+k_{2}+k_{3}+k_{4}+l_{1}+l_{2}+l_{3}+l_{4}+m_{3}+m_{4} + n_{2}+ \frac{43}{18} $, $c'=a'+1$, $a= \sum k_{i} + \sum l_{i} + m_{2}+m_{3}+m_{4} + n_{2} + o+ \frac{19}{6} $, and $c=a+\frac{5}{2}$.


\ 

\section{Discussions}

We have shown that the introduction of the Lauricella hypergeometric functions $F_{D}^{(n)}$ allowed us to calculate disk two-point functions in a wide range of principal $\mathcal{W}$ minimal models. In this section, we discuss the potential application of $F_{D}^{(n)}$ in genus-zero CFT$_2$ correlation function calculations.

Consider CFT$_2$s defined on genus-zero Riemann surfaces with arbitrary numbers of boundaries and crosscaps. In my work \cite{Liu:2023gzf}, I noticed that the genus-$g$, $n$-point, $b$-boundary, $c$-crosscap correlation functions $\mathcal{F}_{g,n,b,c}$ can be expanded as infinite linear combinations of the genus-$g$, $(n+b+c)$-point functions $\mathcal{F}_{g,(n+b+c)}$
\begin{equation}
    \mathcal{F}_{g,n,b,c} = \sum \mathcal{F}_{g,(n+b+c)} . \label{eq: F exp}
\end{equation}
This expansion is valid since the boundary and crosscap states can be expanded as infinite linear combinations of asymptotics state that corresponding to bulk operators. Two technical difficulties exist for applying this method: determining the infinite set of expansion coefficients, and calculating the infinite series of bulk correlation functions. Both difficulties can be solved effectively by applying the free-field approaches.

Next, we apply the Coulomb-gas formalism to the genus-zero bulk correlation functions in the infinite linear combination. The descendant correlation functions are the free-field descendant correlation functions, where we expect that they share the same type of integrand as the primary correlation functions. The primary correlation functions are expected to take the form of 
\begin{equation*}
    \oint dz^{1} \cdots \oint dz^{n} \prod (z^{ij})^{\alpha^{ij}}  (z_{ab})^{\alpha_{ab}} (z^{i}-z_{a})^{ \alpha^{i}_{a}}  
\end{equation*}
\begin{equation}
    \oint d\widebar{z}^{(1)} \cdots \oint d\widebar{z}^{(\widebar{n})} \prod (\widebar{z}^{\widebar{i}\widebar{j}})^{\widebar{\alpha}^{\widebar{i}\widebar{j}}} (\widebar{z}_{\widebar{a}\widebar{b}})^{\widebar{\alpha}_{\widebar{a}\widebar{b}}} ( \widebar{z}^{\widebar{i}} -\widebar{z}_{\widebar{a}} )^{\widebar{\alpha}^{\widebar{i}}_{\widebar{a}}} , 
\end{equation}
which can be obtained analytically by the application of the Taylor expansion and the Pochhammer contour integral of $F_{D}^{(n)}$.

\ 

\appendix

\section{The background charged bosonic free-field approach to semi-simple bosonic QDS $\mathcal{W}$ minimal models}
\label{sec: semi W}

We discuss the rational principle QDS $\mathcal{W}$ minimal models related to bosonic semi-simple KM algebras \cite{Kac:1984mq, Kac:1990gs}. Our discussion focuses on the free-field approach to them, that is, constructing the $\mathcal{W}$ algebra as the centralizer of the screening operators. 

\

Consider a KM algebra $\widehat{g} $ related to a semi-simple Lie algebra $g = g_{1} \oplus g_{2} \oplus\cdots \oplus g_{n} $. Each simple part $\widehat{g}_{i}$ has a central extension, denoted by $k^{(i)}$, and its Coxeter number and dual Coxeter number, denoted as $h^{(i)}$ and $h^{\vee(i)}$ respectively. Analogous to the simple cases, we define parameters $(\alpha_{+}^{(i)})^{-2} = k^{(i)} + h^{\vee (i)}$ and $\alpha_{+}^{(i)}\alpha_{-}^{(i)}=-1$. We denote the simple roots as
\begin{equation}
    \Big\{ \{ \alpha_{i}^{(1)} \} , \{ \alpha_{i}^{(2)}\} , \cdots , \{  \alpha_{i}^{(n)} \} \Big\}, 
\end{equation}
where $i=1,2 , \cdots , r_{1} + r_{2}+ \cdots +r_{n}$, and the superscript labels the corresponding simple parts. The simple roots from different simple parts are in different dimensions in the weight space. The simple coroots and fundamental weights are defined in the same manner as in the simple cases. A semi-simple weight vector $\Lambda$ is denoted by
\begin{equation}
    \Lambda= \big[ \Lambda_{1} , \Lambda_{2}, \cdots , \Lambda_{r_{1}+ \cdots + r_{n}}  \big] = \big(\Lambda^{(1)} , \Lambda^{(2)} , \cdots  , \Lambda^{(n)} \big) .   
\end{equation}

The Wakimoto-type of free-field approach generalizes to semi-simple cases straight-forwardly. Hence, the $\mathcal{W}$ screening operators induced from the QDS procedure take the same form as the simple cases $: e^{ -i \alpha_{\pm} \alpha^{(j)}_{i} \cdot \phi  } : (z)$. The chiral generators of the semi-simple QDS $\mathcal{W}$ algebra are given by
\begin{equation}
    \Big\{  \big[ T^{(1)}(z) , \mathcal{W}^{i(1)} , \cdots  \big] , \cdots ,  \big[ T^{(n)}(z) , \mathcal{W}^{i(n)} , \cdots  \big]  \Big\} , 
\end{equation}
where each block $ \big[ T^{(a)}(z) , \mathcal{W}^{i(a)} , \cdots  \big]$ consist of the free-field simple QDS $\mathcal{W} \big[ \widehat{g}_{a} (k^{(a)}) \big]$ chiral generators. The commutation relations satisfy 
\begin{equation}
    [\mathcal{W}^{i(a)}_{m} ,  \mathcal{W}^{j(b)}_{n}  ] \quad   \propto \quad  \delta^{ab} .  
\end{equation}

A semi-simple $\mathcal{W} \big[ \widehat{g}(k^{(1)} , \cdots , k^{(n)} ) \big]$ Verma module $ V_{\Lambda}$ has a primary vector $\vert v_{\Lambda} \rangle$ satisfying
\begin{equation}
    W^{i(a)}_{n>0} \vert v_{\Lambda} \rangle = 0 , \quad W^{i(a)}_{0} \vert v_{\Lambda} \rangle = w^{i(a)} \vert v_{\Lambda} \rangle, \quad \forall i , a \:. 
\end{equation}
The semi-simple $\mathcal{W}$ Verma modules are tensor products of simple $\mathcal{W}$ Verma modules 
\begin{equation}
    V_{\Lambda} = V_{\Lambda^{(1)}}^{\widehat{g}_{1}} \otimes \cdots \otimes V_{\Lambda^{(n)}}^{\widehat{g}_{n}} . 
\end{equation}
The singular vector structure of $V_{\Lambda}$ is the combination of all the simple $V_{\Lambda^{(i)}}^{\widehat{g}_{i}}$ singular vector structures. The chiral characters of the semi-simple $\mathcal{W}$ modules are the products of simple $\mathcal{W}$ chiral characters 
\begin{equation}
    \text{ch}_{\Lambda} = \text{ch}_{ \Lambda^{(1)}}^{\mathcal{W} [ \widehat{g}_{1}] } \times \text{ch}_{ \Lambda^{(2)}}^{\mathcal{W} [ \widehat{g}_{2}] } \times \cdots \times \text{ch}_{ \Lambda^{(n)}}^{\mathcal{W} [ \widehat{g}_{n}] } .  \label{eq: semi W ch}
\end{equation}
(\ref{eq: semi W ch}) indicates that the semi-simple $\mathcal{W}$ modular matrices are products of the simple $\mathcal{W}$ modular matrices
\begin{equation}
    T_{\Lambda\mu} = \prod_{i=1}^{n} T_{ \Lambda^{(i)}\mu^{(i)}}   , \quad S_{\Lambda\mu}= \prod_{i=1}^{n} S_{ \Lambda^{(i)}\mu^{(i)}}  . \label{eq: mod S semi-sim}
\end{equation}

\

A semi-simple $\mathcal{W} \big[ \widehat{g} (k^{(1)} , k^{(2)}, \cdots , k^{(n)} ) \big]$ algebra is rational when all simple $\mathcal{W} \big[ \widehat{g}_{j}(k^{(j)}) \big]$ are rational. Hence, the semi-simple minimal Kac-table is the combination of all simple Kac-tables
\begin{equation}
    \widehat{\Lambda}^{(+)i} \in \widehat{P}_{+}^{p^{i}-h^{\vee i}}  ,\quad \widehat{\Lambda}^{(-)i} \in \widehat{P}_{+}^{ \vee\:  (p')^{i}-h^{i}} . 
\end{equation}
The Weyl group of a semi-simple Lie algebra is the direct product of the simple Weyl groups
\begin{equation}
    W(g) \cong W(g_{1}) \times W(g_{2}) \times \cdots \times W(g_{n}), 
\end{equation}
hence the expression for the simple $\mathcal{W}$ minimal modular matrix (\ref{eq: mod S conj}) is valid for the semi-simple cases. We apply the Verlinde formula to the rational semi-simple $\mathcal{W}$ minimal models. The relation (\ref{eq: mod S semi-sim}) reveals that the fusion rules of the semi-simple model are the product of the simple $\mathcal{W}$ minimal fusion rules 
\begin{equation}
    N_{ \Lambda \mu }^{\nu} = \prod_{i=1}^{n} N_{ \Lambda^{(i)} \mu^{(i)} }^{\nu^{(i)}} .
\end{equation}

The Fock space vertex operators in the semi-simple $\mathcal{W}$ algebras are 
\begin{equation}
     : e^{ i \Lambda \cdot \phi } : (z)  = \: : e^{ i \Lambda^{(1)}  \cdot \phi^{(1)} } : (z) \times : e^{ i \Lambda^{(2)}  \cdot \phi^{(2)} } : (z) \times \cdots \times : e^{ i \Lambda^{(n)}  \cdot \phi^{(n)} } : (z) .
\end{equation}
The Fock space primary is obtained by 
\begin{equation}
    \vert v_{\Lambda} \rangle =  : e^{ i \Lambda \cdot \phi } : (z)  \vert 0 \rangle = \vert v_{\Lambda^{(1)}} \rangle \otimes  \cdots \otimes  \vert v_{\Lambda^{(n)}} \rangle.
\end{equation}
The free-field resolution of an irreducible minimal module is an $n$-dimensional complex $( F_{\widehat{\Lambda}} ,\: d^{(1)}, \cdots , d^{(n)} )$, where $d^{(a)}$ are simple $\mathcal{W}\big[ \widehat{g}_{a} (k^{(a)}) \big]$ intertwiners. The free-field vertex operators approach to correlation function calculations is analogous to the Coulomb gas formalism to the simple cases. The correlation functions of the semi-simple $\mathcal{W}$ minimal models are again the multiplication of the simple $\mathcal{W}$ minimal parts. This can be shown directly from the integral expression, since the integrals factorize multiplications of independent integrals over simple $\mathcal{W}$ minimal models.

\ 

\section{A Free-field approach to coset $\mathcal{W}$ minimal Ishibashi states}
\label{sec: fre cos}

We show a free-field approach to principal ADE $\mathcal{W}$ coset algebra minimal Ishibashi states. Consider a coset CFT $\widehat{g}(k)/ \widehat{h}(jk)$, $h \subset g$. Its energy-stress tensor is given by 
\begin{equation}
    T(z)= T^{\widehat{g}(k)} (z) -T^{\widehat{h}(jk)} (z) .
\end{equation}
All the coset operators are singular operators of $\widehat{h}(jk)$. A $\widehat{g}(k)$ module $L_{\widehat{\Lambda}}^{\widehat{g}}$ decomposes as the finite direct sum
\begin{equation}
     L_{\widehat{\Lambda}}^{\widehat{g}} =  \bigoplus_{\widehat{\lambda} } \big( L_{(\widehat{\Lambda} , \widehat{\lambda}  )}^{\widehat{g}  /  \widehat{h}}  \otimes L_{\widehat{\lambda}}^{\widehat{h}} \big), \label{eq: coset decom}
\end{equation}
where $L_{(\widehat{\Lambda} , \widehat{\lambda}  )}^{\widehat{g}  /  \widehat{h}}$ are coset modules. Taking the trace over the (\ref{eq: coset decom}), we obtain 
\begin{equation}
   \text{ch}_{\widehat{\Lambda}}  = \sum_{\widehat{\lambda}}  b_{ (\widehat{\Lambda} , \widehat{\lambda}  ) } \times \text{ch}_{\widehat{\lambda}}  , 
\end{equation}
where the coset chiral characters are defined to be the branching functions $b_{ (\widehat{\Lambda} , \widehat{\lambda}  ) }$.

We state the logic of obtaining the coset resolution complex. Assume that a free-field resolution complex of the $\widehat{g}$ module $L_{\widehat{\Lambda}}$ is $(F_{ \widehat{\Lambda} } , d )$. The vectors in the coset modules $L_{(\widehat{\Lambda}, \widehat{\lambda})}^{\widehat{g}/\widehat{h}}$ are $\widehat{h}$ singular vectors with fixed $\widehat{h}$ weight $\widehat{\lambda}$. Hence, the subcomplex $(S_{(\widehat{\Lambda} , \widehat{\lambda})} , d )$ of $(F_{ \widehat{\Lambda} } , d )$ is the resolution complex of coset module $L^{\widehat{g} / \widehat{h} }_{(\widehat{\Lambda} , \widehat{\lambda})}$, where $S_{(\widehat{\Lambda} , \widehat{\lambda})} \subset F_{ \widehat{\Lambda} } $ consists of $\widehat{h}$ singular vectors with fixed $\widehat{\lambda}$ \cite{Bouwknegt:1990fb, Bouwknegt:1990wa}. The universal method of this projection is the BRST procedure, where we project to desired $\widehat{h}$-weight using a $\text{rank}(h)$-dimensional constraint condition. Hence, $\text{rank}(h)$ $bc$ ghost systems are introduced.

\

The coset ADE principal $\mathcal{W} \big[ \widehat{g} \big] $ theories are realized by the diagonal cosets \cite{Goddard:1984vk, Goddard:1986ee, Mathieu:1990dy, Bouwknegt:1992wg}
\begin{equation}
    \widehat{g}^{(1)}(k) \oplus \widehat{g}^{(2)}(1) / \widehat{g}^{(3)}(k+1) ,
\end{equation}
where diagonal means that the KM currents in the three parts satisfy the diagonal embedding $J^{a(3)} (z) =J^{a(1)}(z) +J^{a(2)}(z) $. We denote the diagonal cosets by $\big[ \widehat{g} \oplus \widehat{g} / \widehat{g} , (k,1) \big]$. The coset chiral characters of the ADE diagonal cosets $ \big[ \widehat{g} \oplus \widehat{g} / \widehat{g} , (k,1) \big] $ are identical to that of QDS $\mathcal{W} \big[  \widehat{g} \big] (p,p')$ algebras, at $k+h^{\vee} = p/  (p'-p) := p/u $ \cite{ Bouwknegt:1992wg}. A rigorous proof of the equivalence between simply-laced principal QDS $\mathcal{W}$ algebras and the diagonal coset $\mathcal{W}$ algebra was given in \cite{Arakawa:2018iyk}. The ADE $ \big[ \widehat{g} \oplus \widehat{g} / \widehat{g} , (k,1) \big] $ branching rules are \cite{Bouwknegt:1992wg}
\begin{equation}
    L^{(1)}_{\widehat{\Lambda}^{(1)}} \otimes L^{(2)}_{\widehat{\Lambda}^{(2)}} = \bigoplus\: \big( L^{\mathcal{W}}_{( \Lambda^{(+)} ,\Lambda^{(-)})} \otimes L^{(3)}_{\widehat{\Lambda}^{(3)}} \big) ,
\end{equation}
where $\Lambda^{(1)}+\Lambda^{(2)}-\Lambda^{(3)} \in Q $, $  \widehat{\Lambda}^{(1)} = \widehat{\Lambda}^{(+)} - (u-1) (k'+h^{\vee}) \omega_{0}$ $\widehat{\Lambda}^{(2)} \in \widehat{P}_{+}^{1} $, and $\widehat{\Lambda}^{(3)} = \widehat{\Lambda}^{(-)} -(u-1)(k'+h^{\vee}+1)\omega_{0} $ ($\omega_{0}$ being the zeroth fundamental weight).

The coset resolution has been applied to diagonal ADE coset modules in \cite{Bouwknegt:1990fb}. The projection is realized by the BRST approach, where we introduce $r= \text{rank}(g)$ $bc$ ghost systems. The BRST invariant projection operator is
\begin{equation}
    \widehat{P}_{\Lambda^{(3)}} = \int_{0}^{1} d\theta \:  e^{2\pi i \theta \cdot ( \widehat{h}_{0}-{\Lambda}^{(3)}) }
\end{equation}
where $\widehat{h}_{0}$ is the BRST invariant ($r$-dimensional) Cartan current 
\begin{equation}
  \widehat{h}_{0} =  h_{0}^{(1)} + h_{0}^{(2)} + \text{ghost terms}.   
\end{equation}
The resolution complex of the tensor products of the two dominant modules $L_{  \widehat{\Lambda}^{(1)}}\otimes L_{  \widehat{\Lambda}^{(2)}}$ is the double complex $(  \widetilde{F}_{  \widehat{\Lambda}^{(1)}}  \otimes  \widetilde{F}_{ \widehat{\Lambda}^{(2)}},\:  \widetilde{d}^{(1)} ,\: \widetilde{d}^{(2)} )$, where $[\widetilde{d}^{(1)},\widetilde{d}^{(2)}]=0$. Applying the $H_{Q} \big( F^{bc} \otimes F^{\beta \gamma} \big) \cong \mathbb{C}$ lemma to the $\widetilde{F}_{  \widehat{\Lambda}^{(1)}}  \otimes  \widetilde{F}_{ \widehat{\Lambda}^{(2)}} \otimes F^{bc} $ Fock space modules, we obtain that the Fock space modules in the projected complex are 
\begin{equation}
     F_{  \widehat{w}^{(1)} \ast  \widehat{\Lambda}^{(1)}}^{\phi^{i}}  \otimes F_{  \widehat{w}^{(2)} \ast \widehat{\Lambda}^{(2)}}^{\phi^{i}}  \otimes F^{\beta^{\alpha} \gamma^{\alpha}} .
\end{equation}
The form of the resolution complex are then
\begin{equation}
    \Big[  \widehat{P}_{ w^{(1)} \ast  \Lambda^{(1)} +  w^{(2)} \ast \Lambda^{(2)} - \Lambda^{(3)} } ( F_{  \widehat{w}^{(1)} \ast  \widehat{\Lambda}^{(1)}}^{\phi}  \otimes F_{  \widehat{w}^{(2)} \ast \widehat{\Lambda}^{(2)}}^{\phi}  \otimes F^{\beta^{\alpha} \gamma^{\alpha}} ) \:, \widetilde{d}^{(1)},\: \widetilde{d}^{(2)} \Big] ,
\end{equation}
where the zeroth cohomology space of it is conjectured to be isomorphic to the $\mathcal{W}$ irreducible modules $L_{(\Lambda^{(+)} , \Lambda^{(-)} )}$.

The trivial and charge conjugated gluing conditions of three $\widehat{g}$ algebras transform into the that in the coset model. The trivial and charge conjugated $\widehat{g}$ gluing conditions also transform into that of the $\phi^{i} \beta^{\alpha} \gamma^{\alpha}$ systems. Hence, by applying the free-field resolution conjectures to the coset $\mathcal{W}$ Ishibashi states, we obtain
\begin{equation*}
    \vert  L_{( \Lambda^{(+)} , \Lambda^{(-)} )} \rangle \rangle_{ \Omega }  = \sum_{ \widehat{w}^{(1,2)} }  \widehat{P}_{ w^{(1)} \ast  \Lambda^{(1)} +  w^{(2)} \ast \Lambda^{(2)} - \Lambda^{(3)} } \theta_{\widehat{w}_{1} , \widehat{w}_{2}}  
\end{equation*}
\begin{equation}
    \vert   F_{ \widehat{w}^{(1)} \ast  \widehat{\Lambda}^{(1)}  } \rangle \rangle_{ \Omega } \otimes \vert  F_{ \widehat{w}^{(2)} \ast  \widehat{\Lambda}^{(2)}}  \rangle \rangle_{ \Omega } \otimes \vert F^{\beta^{\alpha} \gamma^{\alpha}}    \rangle \rangle_{I} ,
\end{equation}
where $\Omega = I,U$, and the phases $ \theta_{\widehat{w}_{1} , \widehat{w}_{2}}$ are introduced to produce the correct phase in the overlap. The free-field Ishibashi states $\vert   F_{ \widehat{w}^{(i)} \ast  \widehat{\Lambda}^{(i)}  } \rangle \rangle_{\Omega}$, $i=1,2$, take the form of 
\begin{equation}
    \vert   F_{ \widehat{w}^{(i)} \ast  \widehat{\Lambda}^{(i)}  } \rangle \rangle_{ \Omega } = \exp{ \Big[ \sum_{ n \in \mathbb{Z}_{+} }  \frac{-1 }{n}  a_{-n} \cdot V_{\Omega} \cdot  \widebar{a}_{-n}  \Big] } \vert p \otimes \widebar{p} \rangle  .
\end{equation}
The analysis of the form of the $V_{\Omega}$ and the $p-\widebar{p}$ relation is exactly the same as the $\mathcal{W}\big[ \widehat{g}(k) \big]$ cases. Hence for the diagonal and conjugated cases, $V_{\Omega} = I, U$, and $\widebar{p} = -V_{\Omega}  \cdot ( p +\alpha_{+}\rho ) - \alpha_{+}\rho $. The $\beta^{\alpha} \gamma^{\alpha}$ Ishibashi states take the form of 
\begin{equation}
    \vert F^{\beta^{\alpha} \gamma^{\alpha}}    \rangle \rangle_{I} = \exp{\Big[-  \sum_{ \alpha \in \Delta_{+} } \sum_{n \in \mathbb{Z}_{+} }  (\beta_{-n} \widebar{\gamma}_{-n} + \gamma_{-n} \widebar{\beta}_{-n}) \Big] } \vert 0 \otimes \widebar{0} \rangle,
\end{equation}
\begin{equation}
    \vert F^{\beta^{\alpha} \gamma^{\alpha}}    \rangle \rangle_{U} = \exp{\Big[  \sum_{ \alpha \in \Delta_{+} } \sum_{n \in \mathbb{Z}_{+} }  (\beta_{-n} \widebar{\beta}_{-n} - \gamma_{-n} \widebar{\gamma}_{-n}) \Big] } \vert 0 \otimes \widebar{0} \rangle.
\end{equation}

\

\section{Disk two-point functions in $\mathcal{N}=1$ super-Virasoro minimal model $\mathfrak{SVir}(5,3)$}
\label{sec: N=1}

In this section, we consider the simplest rational supersymmetric $\mathcal{W}$ minimal models, which are the $\mathcal{N}=1$ super-Virasoro minimal models $\mathfrak{SVir}(p,p')$.

A supersymmetric chiral algebra contains both bosonic and fermionic chiral generators. The fermionic chiral generators admit two different monodromy behaviors around the origin of the complex plane $\mathbb{C}$ 
\begin{equation}
    G^{i}(e^{2\pi i } z )  = \pm G^{i} (z). 
\end{equation}
The Laurent expansion of $G^{i} (z)$ is
\begin{equation}
    G^{i} (z) = \sum_{r}  \frac{G_{r}^{i}}{z^{r+h^{i}}}, \quad G_{r}^{i} = \oint \frac{dz}{2\pi i} z^{r+h^{i}-1}  G^{i} (z) .
\end{equation}
For $\mathcal{N}=1$ supersymmetric chiral algebras, there is only one fermionic chiral generator $G(z)$. The algebra is called an NS algebra if $G(e^{2\pi i } z )  = G (z)$. Analogously, the algebra is called an R algebra if $G(e^{2\pi i } z )  = -G (z)$ \cite{Friedan:1984rv, Bershadsky:1985dq}  . The vacuum vectors of the NS and R algebras are denoted by $\vert 0 \rangle$ and $\vert \text{R} \rangle$ respectively. The NS vacuum vector has conformal weight $h=0$, while the R vacuum has conformal weight $h=\frac{1}{16}$ 
\begin{equation}
   \text{NS}: \quad  r \in \mathbb{Z}+ \frac{1}{2} \quad ; \quad  \text{R}:\quad  r \in \mathbb{Z}. 
\end{equation}

The $\mathcal{N}=1$ chiral primary vectors are defined as 
\begin{equation*}
    \text{NS} : \quad  L_{n >0 } \vert v_{i} \rangle = G_{r \ge \frac{1}{2}} \vert v_{i} \rangle =  0 , \quad \forall n ,r \: ; 
\end{equation*}
\begin{equation}
    \text{R} : \quad  L_{n >0 } \vert v_{i} \rangle = G_{r \ge 1} \vert v_{i} \rangle =  0 , \quad \forall n ,r\: . 
\end{equation}

NS and R sector $\mathcal{N}=1$ chiral modules are spanned by different basis
\begin{equation}
    \mathcal{H}_{i}^{\text{NS}}= \text{Span} \: ( G_{-R} L_{-N} \vert v_{i} \rangle )  , R=  r_{}  \ge \frac{1}{2}  , \quad N= \{  \} , \: n_{1} \ge n_{2} \ge \cdots \ge 1
\end{equation}
\begin{equation}
    \mathcal{H}_{i}^{\text{R}}= \text{Span} \: ( G_{-R} L_{-N} \vert v_{i} \rangle ) 
\end{equation}
The R module primary vectors are degenerate, due to the existence of the zeroth mode of the supercurrent $G_{0}$. The degeneracy reduces at certain specific values of the chiral algebras. The simplest example is the $h=c_{\text{Vir}}/24$ point for the $\mathcal{N}=1$ super-Virasoro algebra, where the vector $G_{0} \vert \text{R} \rangle$ is a singular vector.

For each supersymmetric chiral module, two types of chiral chracters can be defined 
\begin{equation*}
     \text{ch}_{L_{i}^{\text{NS}}} := \text{Tr}_{L_{i}^{\text{NS}}}\:  q^{L_{0}-\frac{c}{24}} , \quad    \text{sch}_{L_{i}^{\text{NS}}} := \text{Tr}_{L_{i}^{\text{NS}}}\: (-1)^{F} q^{L_{0}-\frac{c}{24}} , 
\end{equation*}
\begin{equation}
     \text{ch}_{L_{i}^{\text{R}}} := \text{Tr}_{L_{i}^{\text{R}}}\:  q^{L_{0}-\frac{c}{24}} , \quad    \text{sch}_{L_{i}^{\text{R}}} := \text{Tr}_{L_{i}^{\text{R}}}\: (-1)^{F} q^{L_{0}-\frac{c}{24}} , 
\end{equation}
where $F$ is the fermion number operator, counting the number of fermionic modes that act on the bosonic primary vector. The existence of fermionic zero modes ensured that all R-sector super-characters $\text{sch}_{L_{i}^{\text{R}}} $ vanish.

\

The modular matrices of the supersymmetric RCFTs have model-independent properties that can be obtained from the geometrical point of view. For our purpose, we show only the 
\begin{equation}
  S: \quad \text{sch}_{L_{i}^{\text{NS}}} \Longleftrightarrow \text{ch}_{L_{i}^{\text{R}}} ,
\end{equation}
\begin{equation}
    S= \begin{pmatrix}
        S^{\text{NS}, \text{NS}} & 0 & 0 \\ 0 & 0 & S^{\widetilde{\text{NS}} , \text{R}} \\ 0 & S^{\text{R}, \widetilde{\text{NS}}} & 0
    \end{pmatrix} .
\end{equation}
To construction of the modular invariants in the supersymmetric RCFTs is necessary to include three types of the chiral characters. In this work, only the diagonal $\mathcal{N}=1$ super-Virasoro minimal models are considered.

\

The bosonic version of the Verlinde formula does not apply to the supersymmetric models. The generalizations to the supersymmetric cases are pioneered in the works \cite{Mussardo:1987eq, Mussardo:1987ab, Eholzer:1993ek}. The supersymmetric fusion rules can be decomposed into three subclasses $\text{NS} \times \text{NS} \to \text{NS}$, $\text{R} \times \text{R} \to \text{NS}$, and $\text{NS} \times \text{R} \to \text{R}$. The three classes are described by three supersymmetric Verlinde formulas \cite{Eholzer:1993ek}
\begin{equation*}
    \text{NS} \times \text{NS} \to \text{NS} \: : \quad  N_{ij}^{k} = \sum_{l \in \Delta_{\text{NS}} } \frac{S_{il}^{\text{NS},\text{NS}} S_{jl}^{\text{NS},\text{NS}} (S^{\text{NS},\text{NS}}_{kl})^{-1}  }{S_{0l}^{\text{NS},\text{NS}}}  ,
\end{equation*}
\begin{equation*}
    \text{R} \times \text{R} \to \text{NS} \: : \quad  N_{ij}^{k} = d_{i}d_{j} \sum_{l \in \Delta_{\text{NS}} } \frac{S_{il}^{ \text{R};\widetilde{\text{NS}} } S_{jl}^{ \text{R};\widetilde{\text{NS}} } (S_{kl}^{\text{NS};\text{NS}})^{-1} }{S_{0l}} ,
\end{equation*}
\begin{equation}
    \text{NS} \times \text{R} \to \text{R}   \: : \quad  N_{ij}^{k}= d_{i}d_{k}^{-1} \sum_{l \in \Delta_{\text{NS}} } \frac{S^{\text{R};\widetilde{\text{NS}}}_{il} S_{jl}^{\text{NS};\text{NS}} (S_{kl}^{\text{R};\widetilde{\text{NS}}})^{-1} }{S_{0l}^{\text{NS};\text{NS}}}  \label{eq: sup Ver} .
\end{equation}
The factors $d_{i}$ are positive integers that are simple current orbit lengths in the bosonic projection of the supersymmetric algebra $\mathcal{A}$. For our purpose, it is unnecessary to know these $d_{i}$ factors.

\

The $\mathcal{N}=1$ super-Virasoro algebra has the (anti-)commutation relations are 
\begin{equation*}
    [L_{m},L_{n}]=(m-n)L_{m+n}+ \frac{\Hat{c}}{8}(m^3-m)\delta_{m+n,0} , \quad \Hat{c}= \frac{2}{3}c_{Vir} ,
\end{equation*}
\begin{equation}
    [L_{m},G_{r}]= \Big( \frac{m}{2}-r \Big) G_{m+r} , \quad  \{ G_{r} ,G_{s} \} =2 L_{r+s} + \frac{\Hat{c}}{2}\Big( r^{2}-\frac{1}{4} \Big) \delta_{r+s,0} .
\end{equation}

\ 

The rational $\mathcal{N}=1$ super-Virasoro minimal algebra occurs at $\alpha_{+}= p' /p$, leading to the $\mathcal{N}=1$ central charges 
\begin{equation}
    \widehat{c}= 1-\frac{2(p-p')^{2}}{pp'}.
\end{equation}
There are two types of parameterizations \cite{Eholzer:1993ek}
\begin{itemize}
    \item $p,p' \in \mathbb{Z}_{+}$, $\gcd(p,p')=1$, $p+p' \in 2\mathbb{Z}_{+}$;

    \item $p,p' \in 2\mathbb{Z}_{+}$, $\gcd( \frac{p}{2}, \frac{p'}{2} )=1$, $\frac{p}{2}+\frac{p'}{2}\in 2\mathbb{Z}_{+}+1 $. 
\end{itemize}
The $\mathcal{N}=1$ minimal models are unitary only when $\vert p-p' \vert=2$. The Kac table of the $\mathfrak{SVir}^{1}(p,p')$ model is
\begin{equation}
     \big\{ L_{(r,s)} \: \vert \:  1\le r \le p-1, \quad 1\le s\le p'-1  \big\}, 
\end{equation}
where the NS representations satisfy $ r+s \in 2\mathbb{Z}$ and the R representations satisfy $ r+s \in 2\mathbb{Z}+1$. The primary conformal weights are given by 
\begin{equation}
    h_{(r,s)} = \frac{ (p'r -ps)^{2} -(p-p')^2}{8pp'} + \frac{1- (-1)^{r-s} }{32} . 
\end{equation}
The $\mathcal{N}=1$ irreducible modules admit the identification $L_{(r,s)}\equiv L_{(p-r,p'-s)}$. However, we don't claim this to be a Weyl symmetric identification, since the Weyl group of the super finite Lie and KM algebras are not defined in this work.

The structure of the complete degenerate $\mathcal{N}=1$ super-Virasoro modules are identical to those of the complete degenerate Virasoro minimal modules. This hinted that the free-field resolutions of the minimal $\mathcal{N}=1$ super-Virasoro irreducible modules are

The modular $S$ matrices of $\mathcal{N}=1$ super-Virasoro minimal models are similar to those of Virasoro minimal models \cite{Cappelli:1986ed, Matsuo:1986vc}
\begin{equation*}
    S_{(r,s)(r',s')}^{\text{NS};  \text{NS}} = \frac{2}{\sqrt{pp'}} \Big[ \cos{(\frac{2\pi \lambda \lambda' }{4pp'})} - \cos{(\frac{2\pi \widebar{\lambda} \lambda' }{4pp'})} \Big] , 
\end{equation*}
\begin{equation*}
    S_{(r,s)(r',s')}^{\text{R}; \widetilde{\text{NS}}} =  \frac{2}{\sqrt{pp'}} \Big[ \cos{(\frac{2\pi \lambda \lambda' }{4pp'})} - (-1)^{r's'} \cos{(\frac{2\pi \lambda \widebar{\lambda}' }{4pp'})} \Big] ,
\end{equation*}
\begin{equation}
    S_{(r,s)(r',s')}^{ \widetilde{\text{NS}} ;\text{R} } = \frac{1}{\sqrt{pp'}} \big(1+\delta_{r',\frac{p}{2}} \delta_{s',\frac{p'}{2}} \big) \Big[ \cos{(\frac{2\pi \lambda \lambda' }{4pp'})} - \cos{(\frac{2\pi \lambda \widebar{\lambda}' }{4pp'})} \Big]   .
\end{equation}
where $\lambda= p' r-ps$ and $\widebar{\lambda}=p' r+ps$.  Applying the explicit forms of the modular $S$ matrices to the Verlinde formula, the fusion rules of the rational $\mathcal{N}=1$ minimal models can be obtained.

The modular invariants of the $\mathcal{N}=1$ super-Virasoro minimal models also admit the ADE classification as the Virasoro minimal cases \cite{Cappelli:1986ed}. For simplicity, only the diagonal $\mathcal{N}=1$ super-Virasoro minimal models are considered.

\

The rational $\mathcal{N}=1$ super-Virasoro minimal models $\mathfrak{SVir}^{1}(p,p')$ admit a free-field approach by a background charge bosonic field $\phi$ and a free fermionic field $\psi$. The free-field realization of the energy-stress tensor $T(z)$ and the supercurrent are given by
\begin{equation}
    T(z)= - \frac{1}{2} :  \partial \phi   \partial \phi(z)   : - \frac{1}{2} : \psi \partial \psi (z) :  + i  \alpha_{0}^{N=1} \partial^{2} \phi(z)   , \label{eq: N=1 fre T}
\end{equation}
\begin{equation}
    G(z) =  \frac{i}{2} :  \psi \partial \phi (z) : -  \alpha_{0}^{N=1} \partial \psi (z), \label{eq: N=1 fre G}
\end{equation}
respectively. There are two fermionic screening operators $s^{\pm}(z) = \psi \: : e^{- i\alpha_{\pm} \phi } :(z) $ \cite{Ito:1990zi}.

The resolution complex is given by 
\begin{equation}
   C_{r,s}(\Hat{c}_{p,p'}): \quad   \cdots  F_{r,2p+s}^{N=1}   \overset{(Q_{-}^{F})^{s}}{\longrightarrow}  F_{r,-s +2p j }^{N=1}     \overset{(Q_{-}^{F})^{p-s}}{\longrightarrow}    F_{r,s}^{N=1}  \overset{(Q_{-}^{F})^{s}}{\longrightarrow} F_{r,-s}^{N=1}  \overset{(Q_{-}^{F})^{p-s}}{\longrightarrow}  \cdots 
\end{equation}
where we conjecture that irreducible modules in $N=1$ minimal models are isomorphic to the zeroth cohomology space of $C_{r,s}(\Hat{c}_{p,p'})$, with all other cohomology spaces being trivial
\begin{equation}
     H^{i} (C_{r,s}(\Hat{c}_{p,p'})) \cong \begin{cases}
       0 & i \ne 0 ,  \\  \mathcal{H}_{r,s}^{N=1}(\Hat{c}_{p,p'}) , & i= 0. 
    \end{cases} 
\end{equation}

\ 

For boundary diagonal $\mathcal{N}=1$ super-Virasoro minimal models, the gluing conditions are 
\begin{equation}
    T(z) = \widebar{T}(\widebar{z}) , \quad G(z) = \pm \widebar{G}(\widebar{z}) .
\end{equation}
The Ishibashi states correspond to them are denoted by $\vert L_{(r,s)} \rangle \rangle_{\pm}$.

Applying the Fock space reoslution to the Ishibashi states $\vert L_{(r,s)} \rangle \rangle_{\pm}$, we obtain 
\begin{equation}
    \vert L_{(r,s)} \rangle \rangle_{ \pm} = \sum_{j \in \mathbb{Z} } \theta_{j}  \vert F_{ -2pj\pm r ,s} \rangle \rangle_{\pm},
\end{equation}
where the phases $\theta$ are introduced to create the correct $\mathcal{N}=1$ super-Virasoro minimal chiral characters from the overlaps between the free-field Ishibashi states. The free-field Ishibashi states take the form of 
\begin{equation}
    \vert  F_{ -2pj\pm r ,s} \rangle \rangle_{\pm}  = \exp{ \Big[ \sum_{n > 0} \mp \frac{ 1}{n} a_{-n} \widebar{a}_{-n}  \Big] } \exp{ \Big[ \sum_{r }  \pm i \psi_{-r} \widebar{\psi}_{-r}  \Big] }    \vert  p_{ -2pj\pm r ,s} \otimes \widebar{p}_{ -2pj\pm r ,s} \rangle .
\end{equation}
where $\widebar{p}_{ -2pj\pm r ,s} =-p_{ -2pj\pm r ,s} -\alpha_{0} $.

\

To calculate the physical disk correlation functions, we need to consider the form of the supersymmetric Cardy boundary states. The universal supersymmetric states should be constructed based on the form of the supersymmetric Verlinde formula. For specific models, specific supersymmetric Cardy boundary states can be constructed.

\

Next, the calculations of the disk two-point functions using the Coulomb-gas formalisms. In this work, only the $\text{NS}\times  \text{NS}\to \text{NS}$ examples are shown, with calculations of the correlation functions involving the R sector can be obtained from similar manner. In the NS sector, the primary free-field vertex operators are 
\begin{equation}
    V_{(r,s)} (z) =  \: : e^{ i \alpha_{(r,s)}  \phi }  :  (z ) .  
\end{equation}
When there is only one intertwiner inserted, the fermionic field $\psi(z)$ provides a $z^{-\frac{1}{2}}$ factor for the integrand. This property can be shown by considering the following inner product
\begin{equation}
    \langle 0 \vert   \psi(z)   \exp{\Big[  \pm i \sum_{r} \psi_{-r}  \widebar{\psi}_{-r} \Big]}  \vert  0  \rangle  .
\end{equation}
Expanding the exponential and the leading term provides $z^{-\frac{1}{2}}$, with all other terms being $0$ at the limit $z \to 0$. Higher order terms provide higher order in $z$ with operators. Analogous result holds for the anti-holomorphic part.

When there are intertwiners inserted in both the holomorphic and the anti-holomorphic sectors, the fermionic field contribution to the integrand will depend on the fermionic conditions. The results are 
\begin{equation}
    _{+}\langle \langle F_{p} \vert \psi(z) \widebar{\psi} (\widebar{z}) \vert 0 \rangle = (1-  z \widebar{z})^{-1},
\end{equation}
and 
\begin{equation}
    _{-}\langle \langle F_{p} \vert \psi(z) \widebar{\psi} (\widebar{z}) \vert 0 \rangle = (1-  z \widebar{z})^{a},
\end{equation}
respectively. The parameter $a$ remains undetermined. However, this doesn't affect the progression of the calculations.

\

The simplest unitary non-trivial $\mathcal{N}=1$ super-Virasoro minimal model is the $\mathfrak{SVir}(5,3)$ model. The Virasoro central charge of this model is $c_{\text{Vir}}=7/10$. The Virasoro minimal model $\mathfrak{Vir}(5,4)$ is a sub-theory of this model. The Kac-table of this model includes two NS representations $\Delta_{\text{NS}}= \{ L_{(1,1)} \equiv L_{(4,2)}, \: L_{(3,1)} \equiv L_{(2,2)} \}$, and two R representations $\Delta_{\text{R}}= \{ L_{(2,1)} \equiv L_{(3,2)}, \: L_{(4,1)} \equiv L_{(1,2)} \}$. The primary conformal weights of the four presentations are
\begin{equation}
    h_{(1,1)} =0,\quad h_{(3,1)} = \frac{1}{10} , \quad h_{(2,1)} = \frac{3}{80} , \quad h_{(4,1)}= \frac{7}{16} .
\end{equation}
The three blocks in the $\mathfrak{SVir}(5,3)$ modular $S$ matrix are 
\begin{equation}
    S^{\text{NS};\text{NS}} = \begin{pmatrix}
        a & b \\ b & -a 
    \end{pmatrix}  , \quad  S^{\widetilde{\text{NS}};\text{R}} = \begin{pmatrix}
        a & b \\ -b & a
    \end{pmatrix},
\end{equation}
where $a= \sqrt{ \frac{1}{10} (5-\sqrt{5}) }$ and $b= \sqrt{ \frac{1}{10} (5+\sqrt{5}) } $. The $\mathfrak{SVir}(5,3)$ fusion rules are 
\begin{equation*}
    \text{NS} \times \text{NS} \to \text{NS} :\quad  L_{(3,1)} \times L_{(3,1)} = L_{(1,1)} + L_{(3,1)} ,
\end{equation*}
\begin{equation}
    \text{R} \times \text{R} \to \text{NS} :\quad L_{(2,1)} \times L_{(2,1)} = L_{(1,1)} + L_{(3,1)} , \quad L_{(2,1)} \times L_{(4,1)}  = L_{(3,1)} . 
\end{equation}

\

The contributions in the disk correlation functions $\langle b_{(r,s)} \vert \Phi_{(3,1)} (z_{1},\widebar{z}_{1}) \Phi_{(3,1)} (z_{2},\widebar{z}_{2}) \vert 0 \rangle$ are calculated. All correlation functions have a total of four contributions $^{\pm}I_{(3,1)(3,1)}^{(1,1)}:=\:  _{\pm}\langle \langle L_{(1,1)} \vert \Phi_{(3,1)} (z_{1},\widebar{z}_{1}) \Phi_{(3,1)} (z_{2},\widebar{z}_{2}) \vert 0 \rangle $ and $^{\pm}I_{(3,1)(3,1)}^{(3,1)}:=\:  _{\pm}\langle \langle L_{(3,1)} \vert \Phi_{(3,1)} (z_{1},\widebar{z}_{1}) \Phi_{(3,1)} (z_{2},\widebar{z}_{2}) \vert 0 \rangle $.

For contributions $^{\pm}I_{(3,1)(3,1)}^{(1,1)}$, a free-field realization is 
\begin{equation}
    \oint dz^{2} \oint dz^{1} \widebar{V}_{(3,1)} (\widebar{z}_{2}) \widebar{V}_{(3,1)}^{\dag} (\widebar{z}_{1})    V_{(3,1)} (z_{1}) s_{+}(z^{1}) s_{+}(z^{2})V_{(3,1)}(z_{2}) . 
\end{equation}
The disk contributions $^{\pm}I_{(3,1)(3,1)}^{(1,1)}$ are
\begin{equation*}
     \sum_{k_{i} \in \mathbb{N} } \vert z_{1}\vert^{-\frac{2}{5}+2k_{2}}  (1- \vert z_{1} \vert^{2} )^{-\frac{1}{5}}  \frac{(\frac{2}{5})_{\sum_{i} k_{i} } (\frac{3}{5})_{k_{1}} (-\frac{1}{5})_{k_{2}}  }{ (\frac{3}{5})_{\sum_{i} k_{i} } (k_{1})!(k_{2})!}
\end{equation*}
\begin{equation}
   \frac{\Gamma(\frac{2}{5}+\sum_{i}k_{i}) \Gamma^{2}(\frac{2}{5}) \Gamma(\frac{3}{5})}{\Gamma(\frac{4}{5}+\sum_{i}k_{i})} \: _{1}F_{2} \Big[  (\frac{2}{5}+\sum k_{i}), -\frac{1}{5}   ;  (\frac{4}{5}+\sum k_{i} ); \vert z_{1} \vert^{2} \Big].
\end{equation}

A free-field realization of the contributions $^{\pm}I_{(3,1)(3,1)}^{(3,1)}$ is 
\begin{equation}
     \oint d\widebar{z} \oint dz \widebar{V}_{(3,1)}(\widebar{z}_{2}) \widebar{s}^{+} (\widebar{z} )  \widebar{V}_{(3,1)}^{\dag}(\widebar{z}_{1})  V_{(3,1)}(z_{1}) s^{+}(z)V_{(3,1)}(z_{2}) . 
\end{equation}
The disk contribution $^{(+)}I_{(3,1)(3,1)}^{(3,1)}$ is
\begin{equation*}
     \sum_{k\in \mathbb{N}} \vert z_{1} \vert^{ - \frac{1}{5} +2k_{1}+2k_{2} } (1- \vert z_{1} \vert^{2} )^{-\frac{1}{5}}   \frac{(-\frac{1}{10})_{\sum_{i} k_{i} }   (\frac{2}{5})_{k_{1}} (-\frac{1}{5})_{k_{2}}  }{ (\frac{3}{10})_{\sum_{i} k_{i} } (k_{1})! (k_{2})!}
\end{equation*}
\begin{equation}
   \frac{\Gamma(-\frac{1}{10}) \Gamma(\frac{2}{5}) }{\Gamma(\frac{3}{10})}   \frac{ \Gamma(k_{1}-\frac{1}{10}) \Gamma(\frac{6}{5})}{\Gamma(k_{1}+\frac{11}{10})} \: _{1}F_{2} \Big[(k_{1}-\frac{1}{10}) , \frac{3}{5} ; (k_{1}+\frac{11}{10}) ; \vert z_{1}  \vert^{2} \Big].
\end{equation}
The disk contribution $^{(-)}I_{(3,1)(3,1)}^{(3,1)}$ is
\begin{equation*}
     \sum_{k\in \mathbb{N}} \vert z_{1} \vert^{ - \frac{1}{5} +2k_{1}+2k_{2} } (1- \vert z_{1} \vert^{2} )^{-\frac{1}{5}}   \frac{(-\frac{1}{10})_{\sum_{i} k_{i} }   (-\frac{3}{5}-a)_{k_{1}} (-\frac{1}{5})_{k_{2}}  }{ (\frac{3}{10})_{\sum_{i} k_{i} } (k_{1})! (k_{2})!}
\end{equation*}
\begin{equation}
   \frac{\Gamma(-\frac{1}{10}) \Gamma(-\frac{3}{5}-a) }{\Gamma(\frac{3}{10})}   \frac{ \Gamma(k_{1}-\frac{1}{10}) \Gamma(\frac{6}{5})}{\Gamma(k_{1}+\frac{11}{10})} \: _{1}F_{2} \Big[(k_{1}-\frac{1}{10}) , \frac{3}{5} ; (k_{1}+\frac{11}{10}) ; \vert z_{1}  \vert^{2} \Big].
\end{equation}

\

We also consider an $\text{R}\times \text{NS} \to \text{R}$ example $\langle b \vert \Phi_{(3,1)} (z_{1},\widebar{z}_{1}) \Phi_{(2,1)}  (z_{1},\widebar{z}_{1}) \vert 0 \rangle$. In this example, there is only one contribution $I_{(2,1)(3,1)}^{(2,1)}:= \: _{+}\langle \langle L_{(2,1)} \vert \Phi_{(3,1)} (z_{2},\widebar{z}_{2}) \Phi_{(2,1)} (z_{2},\widebar{z}_{2}) \vert 0 \rangle$. A free-field realization of the contribution is 
\begin{equation}
    \oint dz  \: \widebar{V}_{(3,1)} (\widebar{z}_{2})  \widebar{V}^{\dag}_{(2,1)} (\widebar{z}_{1})   s^{+}(z)  V_{(2,1)}(z_{1}) V_{(3,1)}(z_{2}) . 
\end{equation}
The integral result is 
\begin{equation}
    \vert z_{1} \vert^{-\frac{3}{40}} (1-\vert z_{1} \vert^{2})^{-\frac{1}{5}} \frac{\Gamma(\frac{2}{5}) \Gamma(\frac{1}{5}) }{\Gamma(\frac{3}{5})} \: _{1}F_{2} \Big[  \frac{2}{5} , - \frac{1}{5} ; \frac{3}{5} ; \vert z_{1} \vert^{2} \Big]. \label{eq: R N=1 53} 
\end{equation}

\

\acknowledgments

The author would like to thank H.Z.Liang for guidance and encouragement.




\end{document}